\documentstyle[11pt,psfig]{article}
\newlength{\boxedparwidth} 
\newenvironment{boxedtext}%
        {\begin{center} \begin{tabular}{|@{\hspace{.15in}}c@{\hspace{.15in}}|}
        \hline \\ \begin{minipage}[t]{\boxedparwidth}
        \setlength{\parindent}{.25in}}%
{\end{minipage} \\ \\ \hline \end{tabular} \end{center}}
\begin{document}
\renewcommand{\thepage}{\roman{page}}
\thispagestyle{empty}
\hfill {\bf ISRP - TD - 3}\\
\begin{center}
{\bf{\Large MONTE CARLO : BASICS}}
\vskip 75 mm
{\bf K. P. N. Murthy\\[5 mm]
Theoretical Studies Section,\\
Materials Science Division,\\
Indira Gandhi Centre for Atomic Research,\\
Kalpakkam 603 102, Tamil Nadu\\
INDIA\\
e-mail: kpn@igcar.ernet.in}
\vskip 50 mm
Indian Society for Radiation Physics\\
Kalpakkam Chapter\\
January 9, 2000.\\ 
\vskip 10 mm
\end{center}
\begin{boxedtext}
This is a monograph released to mark the occasion of the 
Workshop on " Monte Carlo: Radiation Transport", held
at the Indira Gandhi Centre for Atomic Research (IGCAR),
Kalpakkam, 
from February 7, 2000 to February 18, 2000, sponsored 
by the Safety Research Institute (SRI), Atomic Energy 
Regularity Board (AERB) and jointly organized by the 
Indian Society for Radiation Physics (Kalpakkam Chapter) and
IGCAR, Kalpakkam.
\end{boxedtext}
\vskip 50mm
\begin{boxedtext}
This monograph can be obtained at a nominal cost of  Rs. 50/=
+ Rs. 10/= (handling charges),
from the Head, HASD, 
IGCAR, Kalpakkam 603 102, Tamilnadu, India.
\end{boxedtext}
\newpage
\noindent
{\bf {\Huge Foreword}}
\vskip 10mm

Linearized Boltxmann transport equation is extensively 
used in nuclear engineering to assess the transport 
behaviour of radiation in bulk materials. The solution to this 
equation provides detailed knowledge about the spatial 
and temporal distribution of particle fluc (of neutrons, photons,
{\it etc.} All engineering quantities of practical interest 
such as heat produced, reaction rates, effective multiplication
factors, radiation dose {\it etc.}, are derived therefrom.
This equation is not solvable in closed form except in the 
simplest of the situations. Therefore, numerical
methods of solution are used in almost all applications.
Many standardized computer codes have been developed, and 
validated against data from benchmark experiments.

Numerical techniques of solving the equation also turn out to be
inadequate if the geometry of the problem is complex, as it happens
very often in real life situations. In these instances, Monte Carlo
simulation of the physical processes contained in the equation
is possibly the only way out. This technique, started around late 
1940's, has developed considerably over the years and has now reached 
a high level of sophistication. As a general technique, it finds application
in almost all branches of science as well. 

There is a tendency among the many practitioners of the 
Monte Carlo technique to use the method (particularly the 
readily available computer codes) like a black box, little
realising that care and caution are called for in the 
choice of random number generators, in ensuring sampling 
adequacy, or in interpreting the statistical results 
obtained. There are excellent books dealing with the 
underlying theory of Monte Carlo games; studying the 
books requires some depth in mathematics, scaring away
the beginner. This monograph is intended to present the 
basics of the Monte Carlo method in simple terms. The mathematical
equipment required is not more than college level calculus
and notational familiarity with primitive set theory, All the 
essential ideas required for an appreciation of the Monte Carlo
technique - its merits, pitfalls, limitations - are presented in 
lucid pedagogic style.

Dr. K. P. N. Murthy, the author of this monograph, has well over 
twenty five years of research experience in the application 
of Monte Carlo technique to problem in physical sciences.
he has also taught this subject on a number of occassions and has been 
a source of inspiration to young and budding scientists. 
The conciseness and clarity of the monograph give an indication of his 
mastery overy the subject.

It is my great pleasure, as President of the Indian Society 
for Radiation Physics, Kalpakkam Chapter, to be associated
with the publication of this monograph by the society. Like all
other popular brochures and technical documents brought out earlier,
I am sure, this monograph will be well received.

\hfill A. Natarajan

\hfill {\it President, Indian Society for Radiation Physics}

\hfill {\it (Kalpakkam Chapter)}

\newpage
\noindent
\vspace{50mm}

\hfill {\bf {\Huge {\cal Preface}}}

\bigskip
\bigskip
\noindent
Monte Carlo is a powerful numerical technique useful for solving several complex problems. 
The method has gained in importance and popularity owing to the easy availability of 
high-speed computers.

In this monograph  I shall make an attempt to present the theoretical basis for the 
Monte Carlo method in very simple terms: sample space, events, probability of  events, 
random variables, mean, variance, covariance, characteristic function, moments, 
cumulants, Chebyshev inequality, law of large numbers, central limit theorem, 
generalization of the central limit theorem through L\'evy stable law, random numbers, 
generation of pseudo random numbers, randomness tests, random sampling techniques: 
inversion, rejection and Metropolis rejection; sampling from a Gaussian, analogue 
Monte Carlo, variance reduction techniques with reference to importance sampling, 
and  optimization of importance sampling.  I have included twenty-one assignments, 
which are given as boxed items at appropriate places in the text. 

While learning Monte Carlo, I benefited from discussions with many of my colleagues.  
Some are P.~S.~Nagarajan, M.~A.~Prasad, S.~R.~Dwivedi, P. K. Sarkar, 
C.~R.~Gopalakrishnan, M.~C.~Valsakumar, T.~M.~John, R.~Indira, and V.~Sridhar.  
I thank all of them and  several others not mentioned here.

I thank V.~Sridhar for his exceedingly skillful and enthusiastic support to this 
project, in terms of critical reading of the manuscript, checking explicitly the 
derivations, correcting  the manuscript in several places to improve its 
readability and for several hours of discussions.

I thank M.~C.~Valsakumar for sharing his time and wisdom; for his imagination, 
sensitivity and robust intellect which have helped me in my research in general 
and in this venture in particular.  Indeed M.~C.~Valsakumar has been and shall 
always remain a constant source of inspiration to all my endeavours.
 
I thank R. Indira for several hours of discussions and for a critical reading 
of the manuscript.

The first draft of this monograph  was prepared on the basis of the talks  I gave at the 
Workshop  on Criticality calculations using KENO, held at Kalpakkam during 
September 7-18, 1998. I thank A.~Natarajan and C. R. Gopalakrishnan for the invitation.

Subsequently, I spent a month from October 5, 1998, at the School of
Physics, University of Hyderabad, as a UGC visiting  fellow. During this period I gave a 
course on \lq Monte Carlo: Theory and Practice\rq . This course was given in two parts. 
In the first part, I covered the basics and in the second part I discussed several applications.  
This monograph  is essentially based on the first part of the Hyderabad course. I thank 
A.~K.~Bhatnagar, for the invitation. I thank  V.~S.~S.~Sastri,  K.~Venu and  their students 
for the hospitality.

This monograph in the present form was prepared after several additions and extensive revisions 
I made during my stay as a guest scientist at the Instit\"ut f\"ur Festk\"orperforschung, 
Forschungszentrum J\"ulich, for three months starting from 22 July 1999.  I thank  Klaus W. Kehr 
for the invitation. I thank Forschungszentrum J\"ulich for the hospitality. 

I thank Klaus~W.~Kehr, Michael~Krenzlin, Kiaresch Mussawisade, Ralf Sambeth, Karl-Heinz Herrmann, 
D. Basu, M. Vijayalakshmi and several others, for the wonderful time I had in J\"ulich.

I thank Awadesh Mani, Michael Krenzlin, Ralf Sambeth, Achille Giacometti, S. Rajasekar  
Subodh R. Shenoy and S. Kanmani for a critical reading of the manuscript  and for suggesting several changes 
to improve its   readability.

I owe a  special word of thanks to A. Natarajan; he was instrumental in my taking up this project.  
He not only encouraged me into  writing  this monograph but also undertook the responsibility of 
getting it published through the Indian Society for Radiation Physics (ISRP), Kalpakkam Chapter,
on the occassion of the Workshop on Monte Carlo: Radiation Transport, February 7 - 18, 2000, 
at Kalpakkam, conducted by the Safety Research Institute of Atomic Energy Regulatory
Board (AERB).   I am very pleased that 
A. Natarajan has written the foreword to this  monograph.

I have great pleasure in dedicating this monograph to two of the wonderful scientists 
I have met and whom I hold in  very high esteem: {\bf Prof. Dr.  Klaus W.  Kehr}, J\"ulich, who 
retired formally in July 1999, and {\bf Dr. M. A. Prasad}, Mumbai, who is retiring formally in 
March 2000. I take this opportunity to wish them  both the very best.
\\[10mm]
\noindent
Kalpakkam,\\
\noindent
January 9,  2000. \hfill K. P. N. Murthy 
\newpage
\thispagestyle{empty} 
\vskip 5cm
\hfill {\Large  TO}
\vskip 3cm
\hfill {\Huge {\bf  Klaus W. Kehr}},

\bigskip

\hfill {\Large and}

\bigskip

\hfill {\Huge  {\bf M. A. Prasad}}
\newpage

\tableofcontents

\newpage

\pagenumbering{arabic}
\setcounter{page}{1}

\noindent 
\section{INTRODUCTION}

\noindent
Monte Carlo is a powerful numerical technique that makes 
use of random numbers to solve a problem. 
I assume we all know what random numbers are.
 However this issue is, by no means, trivial and  I 
shall have something to say on this later.

Historically, the first large scale Monte Carlo work carried out dates 
back to the middle of the twentieth century. 
This work pertained to studies of neutron multiplication,  scattering, propagation and eventual 
absorption in a medium or leakage from it. Ulam, von Neumann and Fermi were the first to propose  
and employ the  Monte Carlo  method as a viable numerical technique for solving practical problems.

There were of course several isolated and perhaps not fully developed instances  earlier, when  
Monte Carlo has been used in some form or the other.  An example is the experiment performed in 
the middle of the nineteenth century, consisting of throwing a needle randomly on a board notched 
with parallel lines, and inferring the value of $\pi$ from the number of times the needle intersects  
a line; this is  known as Buffon's needle problem,  see for example \cite{HALL}.
 
The Quincunx constructed by Galton \cite{GALTON} toward the end of the nineteenth  century, consisted 
of balls rolling down an array of pins (which deflect the balls randomly to their left or right) 
and getting collected in the vertical compartments placed at the bottom. The heights of the balls in 
the compartments approximate the binomial distribution. This Monte Carlo experiment is a simple 
demonstration of the Central Limit Theorem.

In the nineteen twenties, Karl Pearson perceived the use of random numbers for solving complex 
problems in probability theory and statistics that  were not amenable to exact solutions. Pearson 
encouraged L. H. C. Tippet to produce a table of random numbers to help in such studies, and a book 
of random sampling numbers \cite{lhctippet} was published in the year 1927. This was followed by 
another publication of random numbers  by R. A. Fisher and F. Yates.  Pearson and his students used 
this method to obtain the distributions of several complex statistics.

In India, P. C. Mahalanbois \cite{pcm} exploited \lq\ random sampling \rq\ technique to solve 
a variety problems like the choice of optimum sampling plans in survey work, choice of optimum size 
and shape of plots in experimental work {\it etc.}, see \cite{crrao}.

Indeed,  descriptions of several modern Monte Carlo techniques appear in a paper by 
Kelvin \cite{KELVIN}, written nearly hundred  years ago, in the context of a discussion on the 
Boltzmann equation. But Kelvin was more interested in the results  than  in the technique, 
which to him was {\it obvious}!

Monte Carlo technique derives its name from a game very popular in Monaco. The children 
get together at the beach and   throw pebbles at random on a square which has a circle 
inscribed in it.  From the fraction of the pebbles that fall inside the circle, one can 
estimate the value of $\pi$. This way of estimating the value of $\pi$ goes under the name 
rejection technique.  A delightful variant of the game is discussed by Krauth \cite{krauth}; 
this variant brings out in a simple way the essential principle  behind the Metropolis 
rejection method. We shall discuss the rejection technique as well as the Metropolis 
rejection technique later.

In this monograph I shall define certain minimal statistical terms and invoke some important 
results in mathematical statistics to lay a foundation for the Monte Carlo methods. I shall 
try to make the presentation as simple and as complete as possible.

\noindent
\section{SAMPLE SPACE, EVENTS AND PROBABILITIES}

\noindent
Consider an experiment, real or imagined, which leads to more than one outcome. 
We collect all the  possible outcomes of the experiment in a set $\Omega$ and 
call it the {\bf sample space}. An outcome is often denoted by the symbol $\omega$. 
Certain subsets of $\Omega$ are called {\bf events}. The class of all events is 
denoted by ${\cal F}$.  For every pair of events ${\cal A}_1\in {\cal F}$ and 
${\cal A}_2\in {\cal F}$,  if ${\cal A}_1\cup {\cal A}_2$, ${\cal A}_1\cap {\cal A}_2$, 
and ${\cal A}_1 -{\cal A}_2$ are also events $\in {\cal F}$, then ${\cal F}$ is 
called a {\bf field}.  To each event ${\cal A}$ we assign a real number $0\le {\cal
P}({\cal A})\le 1$ called the {\bf probability} of the event.  If $\Omega$ consists 
of infinitely many  outcomes, we demand that $\cup_{i=1}^{\infty}
{\cal A}_i$ and $\cap_{i=1}^{\infty}{\cal A}_i $ be also events and
form a {\bf Borel field} of all possible events which includes the null
event $\phi$.  We have ${\cal P}(\phi)=0$, ${\cal P}(\Omega) =1$ and
${\cal P}( {\cal A}_1 \cup {\cal A}_2 )={\cal P} ( {\cal A}_1 ) + {\cal
P}({\cal A}_2 )-{\cal P} ( {\cal A}_1
\cap {\cal A}_2)$. Events ${\cal A}_1$ and ${\cal A}_2$ are {\bf disjoint}  
(or {\bf mutually exclusive}) if ${\cal A}_1 \cap {\cal A}_2 = \phi$.

Not all the subsets of $\Omega $ shall be events. One reason for this is that we may wish 
to assign probabilities to only some of the subsets. Another reason is of mathematical 
nature: we may not be able to assign probabilities to some of the subsets of $\Omega$ at all. 
The  distinction between subsets of $\Omega$ and events and the consequent concept of Borel 
field are stated here simply for completeness. In applications, all {\it reasonably defined} 
subsets of $\Omega$ would be events. \\

\noindent
{\bf  How do we attach probabilities to the events~?}\\
 
\noindent
Following the classical definition, we say that ${\cal P}( {\cal A})$ is the ratio of the number of outcomes in the event ${\cal A}$ to that in $\Omega$, provided all {\it the outcomes are equally likely}. To  think of it, this is the method we adopt intuitively to assign  probabilities - like when we say the probability for heads in a toss of a coin is half; the probability for the number, say one, to show up in a roll of a die is one-sixth; or when we say that all micro states are equally probable while formulating  statistical mechanics.  An  interesting problem in this context is the Bertrand paradox  \cite{bertrand}; you are asked to find the probability for a randomly chosen chord in a circle of radius  $r$ to have a length exceeding  $r\sqrt{3}$. You can get three answers $1/2, 1/3,$ and $1/4$, depending upon the experiment you design to select a chord randomly. An excellent discussion of the Bertrand paradox can be found in \cite{PAP}.

\begin{boxedtext}
\noindent
{\bf Assignment 1}\\

\noindent
(A) A fine needle of length $2a$ is dropped at random on a 
board covered with parallel lines with distance $2b$ apart. 
Show that the probability that the needle intersects one of 
the lines equals $2a/\pi b$.  See page 131 of \cite{PAP}.\\

\noindent
(B) Devise an experiment to select randomly a chord in a
 circle of radius $r$. What is the probability that its 
length exceeds $r\sqrt{3}$~?  See page 9-10 of \cite{PAP}. 
What is the probability distribution of the chord length ? \\
\end{boxedtext}

\noindent
Alternately, we can take an operational approach.  ${\cal P}({\cal A})$ is obtained by observing the  frequency of occurrence of ${\cal A}$: repeat the experiment some $N$ times and let $N_{ {\cal A}}$ be the number of times the event ${\cal A}$ occurs.  Then $N_{ {\cal A}}/N$  in the limit of $N\to\infty$ gives the probability of the event.

As seen above, the formal study of probability theory requires three distinct notions namely the sample space $\Omega$, the (Borel) field ${\cal F}$ and the probability measure ${\cal P}$. See Papoulis \cite{PAP} and Feller \cite{FEL}.  The physicists however, use a different but single notion, namely the {\bf ensemble}.  Consider for example a sample space that contains discrete outcomes.  An ensemble is a collection whose members are the elements of the sample space but repeated as many times as would reflect their  probabilities.  Every member of $\Omega$ finds a place in the ensemble and every member   of the ensemble is some element of the sample space. The number of times a given element (outcome) of the sample space  occurs in the ensemble is such that the ratio of  this number to the total number of members in the ensemble is, exactly, the probability associated with the  outcome. The number of elements in the ensemble is strictly infinity.
  
The molecules of a gas in equilibrium is a simple example of an ensemble; the speeds of the molecules at any instant of time have Maxwell-Boltzmann distribution, denoted by $p(v)$.  Each molecule can then be thought of as a member of the ensemble; the number of molecules with speeds between $v_1$ and $v_2 $, divided by the total number of molecules $N$ in the gas is $\int_{v_1}^{v_2} p(v)dv$.

\noindent
\section{RANDOM VARIABLES}

\noindent
The next important concept in the theory of probability is the definition of a (real) {\bf random variable}. A random variable, denoted by $X(\omega)$, is a set function: it attaches a real number $x$ to an outcome $\omega$. A random variable thus maps the abstract outcomes to the numbers on the real line. It stamps each outcome, so to say, with a real number.

Consider an experiment whose outcomes are {\bf discrete} and are denoted by  $\{ \omega_i\}$, with $i$ running from $1$ to say $N$.  Let $X(\omega)$ be the random variable that maps the abstract outcomes $\{ \omega _i \}$ to  real numbers $\{ x_i \}$.  Then $p_i ={\cal P}(\omega\vert X(\omega)=x_i)$ gives the probability that the random variable $X(\omega )$ takes a real value $x_i$.  The probabilities  $\{ p_i \}$ obey the conditions,
\begin{eqnarray}
0\ \    \le\ \  p_i \ \   &
\le \ \  & 1,\  \   \forall\  \ i ,\nonumber\\
  \sum_{i=1}^{N}  p_{i}^{} & =\ \ 1.
\end{eqnarray}
\noindent
Let me illustrate the above ideas through a few examples. \\

\noindent
{\bf Tossing of a single coin}\\

\noindent
The simplest example is the tossing  of a single  coin. The sample space is $\Omega = \{ H,T\} $, where $H$  denotes the outcome Heads and T the Tails. There are four possible events\ :
\begin{eqnarray}
{\cal F}&=&\left\{  \left\{ H \right\}, \left\{ T \right\},
\left\{ H, T \right\}, \left\{ \phi\right\} \right\} .\nonumber
\end{eqnarray} 
The corresponding probabilities are $1/2,\  1/2,\  1$ and $0$ for a fair coin.

We can define a random variable $X(\omega)$ by attaching $+1$ to H  and $-1$ to T, see Table (1).  
Then we say that the probability for the random variable $X$ to take a value $+1$ is ${\cal P} [ \omega\vert X (\omega )=1]=1/2$ and that for $-1$ is half. This defines the discrete probability distribution: $p(1)=1/2,\  p(-1)=1/2$, see Table (1). 

\setcounter{table}{0}
\begin{table}[h]\label{onetoss_tab}
\caption{\protect\small  Random variable and probability for the
 toss of a single coin.}
\bigskip
\begin{center}
\begin{tabular}{|c|c|c|}
\hline
\hspace{25mm}    & \hspace{25mm}  &\hspace{25mm} \\ 
    $\omega$\   & $x= X(\omega)$  & ${\cal  P} \left[ \omega\vert
X(\omega)=x\right]$\\ &               &   \\
\hline
           &                &   \\ H  &    $+1$       & $ 1/2$\\ T  &   $
-1$       & $ 1/2$\\ &               &        \\
\hline
\end{tabular} 
\end{center}
\end{table}

The physicists' ensemble containing ${\cal N}$ members  would be such that half of ${\cal N}$ are  Heads and half Tails.  If the probability of Heads is  $3/4$ and of Tails is $1/4$, then the  corresponding ensemble would contain $3{\cal N}/4$ heads and ${\cal N}/4$ Tails.\\

\noindent
{\bf Tossing of several coins}\\

\noindent
Consider now the tossing of $N$ independent fair coins. Let $X_1$, $X_2$, $\cdots$ $X_N$ be the corresponding random variables.  $X_i (H) = +1$ and $X_i (T) = -1\ \ \forall\ i=1,2,\cdots , N$.  For each coin there are two possible (equally likely) outcomes. Therefore, for $N$ coins there are $2^N$ possible (equally likely) outcomes. For the $N$-coin-tossing experiment each outcome is a distinct string of $H$ and $T$, the string length being $N$.

An example with $N=3$ is shown in Table (2),  
where the random variable $Y_3 (\omega)$ is defined by the sum: $Y_3  = X_1 + X_2 +X_3 $.

\begin{table}[h]\label{sum3rv_tab}
\caption{\protect\small  Random variable and Probability for the  three - coin -tossing experiment}
\bigskip
\begin{center}
\begin{tabular}{|c|c|c|}
\hline
\hspace{25mm} &\hspace{25mm}     &\hspace{25mm}    \\
$\omega$      &  $x=Y_3 (\omega)$  & ${\cal P}\left[ \omega\vert Y_3
(\omega)=x\right] $           \\ &                  &             \\
\hline
              &                  &             \\ HHH     &  +3
& $1/8$       \\ &                  &             \\
\hline
              &                  &             \\ HHT         &
&             \\ HTH         &  $+1$            & $3/8$        \\ THH
&                  &             \\ &                  &             \\
\hline
              &                  &              \\ TTH         &
&        \\ THT         &  $-1$            &       $3/8$  \\ HTT         &
&      \\ &                  &          \\
\hline
              &                  &  \\ TTT          &  $-3$            &
$1/8$  \\ &                  &          \\
\hline
\end{tabular}
\end{center}
\end{table} 
 
\begin{figure}[ht]
\centerline{\psfig{figure=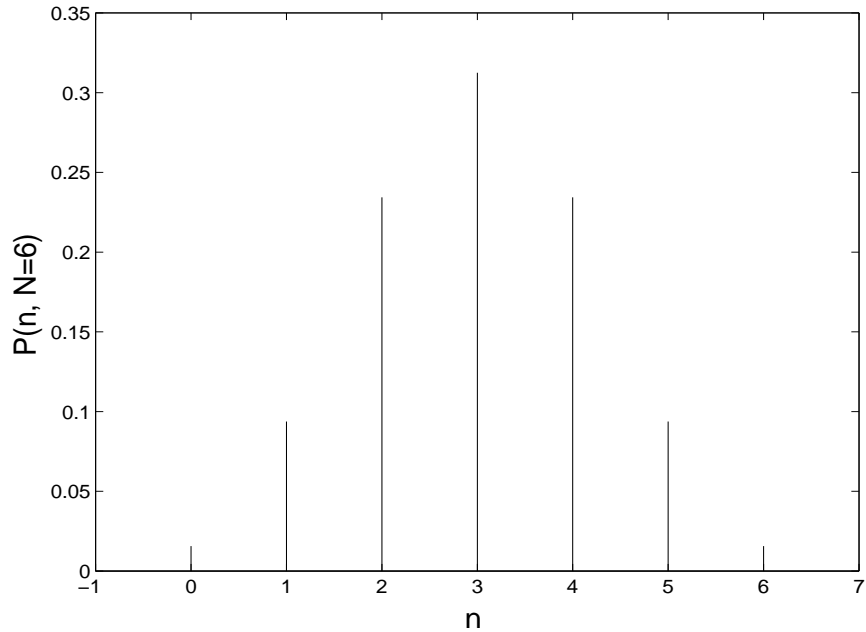,height=8.5cm,width=11.5cm}}
\caption{\protect\small  Binomial distribution, 
see Eq. (\ref{binomial_eq}) with $N=6$  }
\label{BINOMIAL_PS}
\end{figure}

The probability for each outcome (string) in the throw of $N$ coins is thus $2^{-N}$.  Let us define a random variable $Y_N =X_1 + X_2 + \cdots + X_N$.  The probability for $n$ heads in a toss of $N$ coins, which is the same as the probability for the random variable $Y_N $ to take a value $n-(N-n)$ is given by the {\bf binomial} distribution, 
\begin{equation}\label{binomial_eq}
P(n, N) \equiv {\cal P}[Y_N =2n-N]={{1}\over{2^N}}\  {{N!}\over{n! (N-n)!}}.
\end{equation}
Fig. \ref{BINOMIAL_PS} depicts the binomial distribution for $N=6$.

For the general case of a loaded coin, with probability for Heads as $p$ and for the Tails as $q=1-p$, we have,
\begin{equation}\label{loadbin}
P(n,N)\equiv {\cal P}[Y_N =2n-N]=  {{N!}\over{n!\  (N-n)!}}\ p^n\  q^{N-n}.
\end{equation}
\newpage
\noindent
{\bf Rolling of a fair die}\\

\noindent
Another simple example consists of rolling  a die.  There are six possible outcomes. The sample space is given by,
\begin{eqnarray}
\Omega =\hspace{10cm}\nonumber  
\end{eqnarray}
\begin{tiny}
\begin{eqnarray} 
\left\{ 
\begin{array}{cccccccccccccccccccccc}
   &  &  &\bullet&  &  &  &\bullet&  &  &  &\bullet&  &\bullet &
&\bullet&  &\bullet&  &\bullet&\bullet&\bullet  \\ \bullet & & & & & & &
&\bullet & & & & & & & &\bullet & & & & & \\ & &, & & &\bullet &, & &
&\bullet &, &\bullet & &\bullet &, &\bullet & &\bullet &, &\bullet
&\bullet &\bullet
\end{array}
\right\} .\nonumber
\end{eqnarray}
\end{tiny}
The random variable $X(\omega)$ attaches the numbers $\{ x_i : i=1,2,\cdots ,6\}$ $
\equiv  $ $ \{1,\ 2,\ 3,\ 4,\ 5,\ 6\}$ to the outcomes $\{ \omega _i :i=1,2,\cdots ,6\}$, in the same order. If the die is not loaded we have $p(x_i)=1/6\  \forall\  i=1,2,\cdots ,6$.
An example of an event, which is the subset of $\Omega$, is
\begin{eqnarray}
 {\cal A}& =&
\begin{tiny}
 \left\{
\begin{array}{ccccccccc}
 & &\bullet & & & & \bullet & &\bullet \\
\bullet & & &\bullet & & & &\bullet & \\
 &, & & &\bullet &, &\bullet & &\bullet
\end{array}
 \right\} ,\nonumber
\end{tiny}
\end{eqnarray}
 and ${\cal P}({\cal A})=1/2$.  This event corresponds to the roll of  an odd number.  Consider the event
\begin{eqnarray}
{\cal B}=
\begin{tiny}
\left\{
\begin{array}{ccccccccccc}
\bullet & & & &\bullet & &\bullet & &\bullet &\bullet & \bullet\\
        & & & &        & &        & &        &        &         \\ & &
\bullet &, &\bullet & &\bullet &, & \bullet &\bullet&\bullet
\end{array}
\right\} ,
 \nonumber
\end{tiny}
\end{eqnarray}
 which corresponds to the roll of an even number. 
It is clear that the events ${\cal A}$ and ${\cal B}$ 
can not happen simultaneously; in other words a single 
roll of the die can not lead to both the events 
${\cal A}$ and ${\cal B}$. The events ${\cal A}$ 
and ${\cal B}$ are {\bf disjoint} (or {\bf mutually exclusive}.   
If we define another event
\begin{eqnarray}
{\cal C} & = &
\begin{tiny}
 \left\{
\begin{array}{ccccccccc}
 & &\bullet & & & &\bullet & & \\
\bullet & & & & & & &\bullet & \\
 &, & & &\bullet &, & & &\bullet
\end{array}
\right\} ,
\nonumber
\end{tiny}
\end{eqnarray}
then it is clear that ${\cal A}$ and ${\cal C}$ are not disjoint.  We have
\begin{eqnarray}
{\cal A} \cap {\cal C}=
\begin{tiny}
\left\{ 
\begin{array}{ccccc}
 &  & \bullet & & \\
\bullet &  & &\bullet &  \\
 & , &         & &\bullet
\end{array}
\right\} .
\nonumber
\end{tiny}
\end{eqnarray}
Similarly the events ${\cal B} $ and ${\cal C}$ are not disjoint:
\begin{eqnarray}
{\cal B} \cap {\cal C}=
\begin{tiny}
\left\{ 
\begin{array}{ccc}
\bullet & & \\
        & &       \\ & &\bullet
\end{array}
\right\} .\nonumber
\end{tiny}
\end{eqnarray}
\begin{boxedtext}
\noindent
{\bf Assignment 2}\\

\noindent
Consider rolling  of a fair die $N$ times. Let $n_k $ denote the number of times the number $k$ shows up and $k$ runs from $1$ to $6$. Find an expression for the probability $P(n_1 , n_2 , n_3 , n_4 , n_5 , n_6 , N)$. What is the probability distribution of the random variable $n=\sum_{k=1}^{6} k\ n_k$?
\end{boxedtext}

\noindent
{\bf Poisson distribution}\\

\noindent
An important discrete distribution, called the  {\bf Poisson} distribution, arises in the context of random occurrences in time. In an interval $\Delta t\to 0$ positioned at any time $t$, there is either one occurrence with probability $\lambda\Delta t$, or no occurrence with probability $1-\lambda\Delta t$. Here $\lambda^{-1}$ is the characteristic time constant of the Poisson process.

\begin{figure}[ht]
\centerline{\psfig{figure=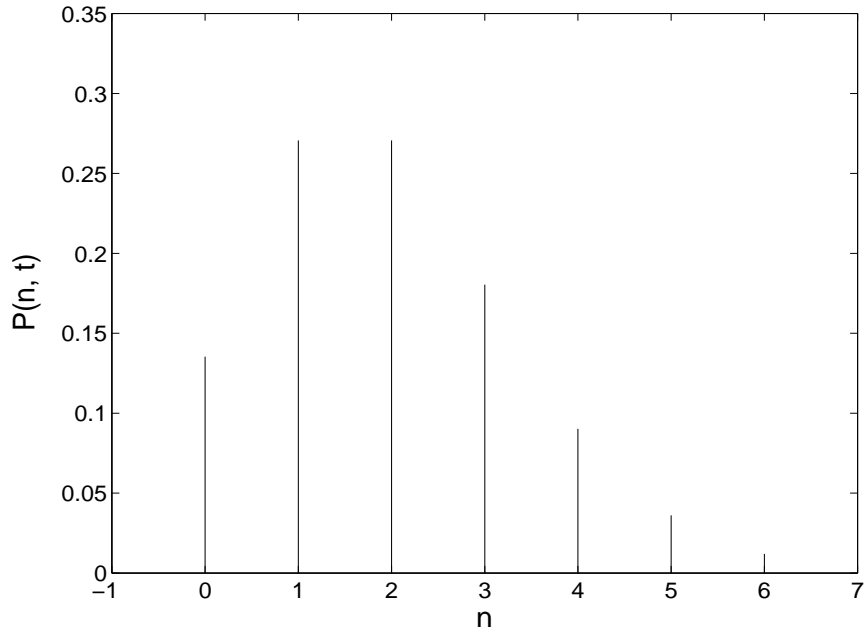,height=8.5cm,width=11.5cm}}
\caption{\protect\small 
Poisson distribution, see Eq. (\ref{poissondist}) with $\lambda t = 2$.}
\label{POISSON_PS}
\end{figure}

Let $P(n,t)$ denote the probability that there are $n$ occurrences in the interval $[0,t]$. A {\bf  master equation} can be readily written down as, \begin{eqnarray}\label{MEP}
P(n,t) & = & \lambda\Delta t\  P(n-1,t-\Delta t)  + [ 1-\lambda\Delta t ]
P(n, t-\Delta t), \nonumber\\ P(n,t=0)&=&\delta_{n,0}.
\end{eqnarray} 
To solve for $P(n,t)$, define the {\bf generating function},
\begin{equation}
\tilde{P}(z,t) = \sum_{n=0}^{\infty}z^n P(n,t).
\end{equation}
Multiplying both sides of Eq. (\ref{MEP}) by $z^n$ and summing over $n$
from $0$ to $\infty$, we get,
\begin{eqnarray}
\tilde{P}(z,t)&=&\lambda z\Delta t\tilde{P} (z,t-\Delta t)
      +  [ 1-\lambda\Delta t ]
\tilde{P}(z, t-\Delta t),\nonumber\\
\tilde{P}(z,t=0)&=& 1.
\end{eqnarray}
 In the limit $\Delta t\to 0$, we get from the above
\begin{equation}
{{ \partial{\tilde{P } }}\over{\partial{t} } } =- \lambda
(1-z)\tilde{P}(z,t),
\end{equation}
whose solution is,
\begin{equation}\label{poigen}
\tilde{P}(z,t)=\exp\left[ -\lambda\left( 1-z\right) t\right],
\end{equation}
since the initial condition,
\begin{eqnarray}
\tilde{P}(z,t=0) &=&\sum_{n=0}^{\infty}z^n P(n,t=0)\nonumber\\
                 & &                                \nonumber\\
&=&\sum_{n=0}^{\infty}z^n \delta _{n,0}\nonumber\\ & &
\nonumber\\ &=&1.
\end{eqnarray}
Taylor expanding the right hand side of Eq. (\ref{poigen}), we get
$P(n,t)$ as the coefficient of  $z^n$,
\begin{equation}\label{poissondist}
P(n,t)={{ (\lambda t)^n}\over{n!}} \exp (-\lambda t) .
\end{equation}
Fig. \ref{POISSON_PS} depicts the Poisson distribution.\\

\noindent
{\bf Continuous distributions}
\\

\noindent
What we have seen above are a few examples of discrete random variable that can take either a finite number or at most countable number of possible values. Often we have to consider random variables that can take values in an interval.  The sample space is  a {\bf continuum}. Then we say that $f(x)dx$ is the probability of the event for which the random variable $X(\omega)$ takes a value between $x$ and $x+dx$. {\it i.e.} $f(x)dx={\cal P}[\omega\vert x\le X(\omega) \le x+dx]$.  We call $f(x)$ the {\bf probability density function} or the {\bf probability distribution function}. \\

\noindent
{\bf Uniform distribution}\\

\noindent
An example is the {\bf uniform} random variable, $U(a,b)$, defined in the interval $[a,b]$. The  probability density function, $f(x)$  of the random variable $U(a,b)$ is defined by $f(x)dx=dx/(b-a)$.  We shall come across the random variable $U(0,1)$ later in the context of pseudo random numbers and their generators.
\\

\begin{figure}[ht]
\centerline{\psfig{figure=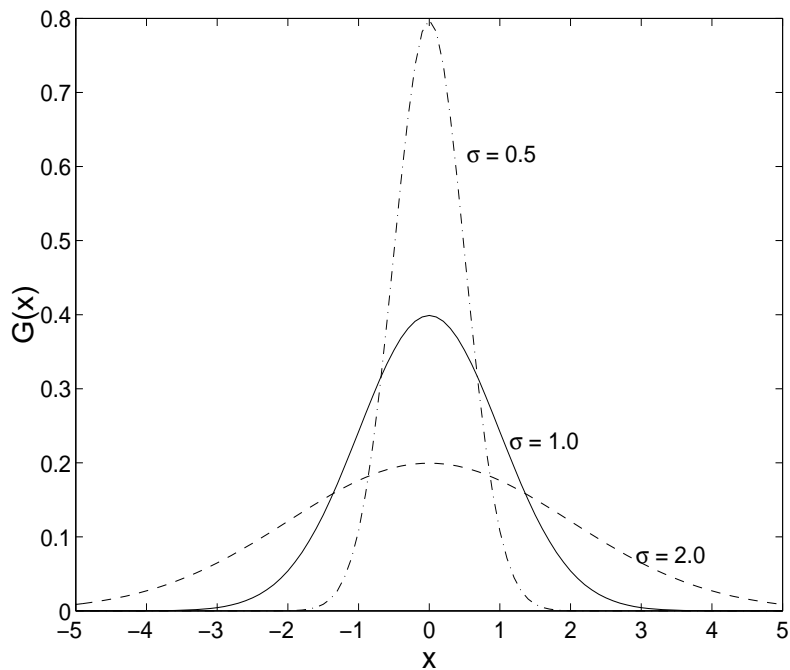,height=9.0cm,width=10.5cm}}
\caption{\protect\small 
Gaussian distribution, see Eq. (\ref{gauss_eq})  with $\mu =0$, and
$\sigma = .5$ (-$\cdot$-$\cdot$), $1.0$ (-)    and $2.0$ (-\  -)}
\label{GAUSS_PS}
\end{figure}

\noindent
{\bf Gaussian distribution}\\

\noindent
A  probability density function we  would often come across is the {\bf Gaussian}.  This density function is of fundamental importance in several physical and mathematical applications.  It is given by, 
\begin{equation}\label{gauss_eq}
G(x)={{1}\over{\sigma \sqrt{ 2\pi}}}\exp\left[- {{ \left( x-\mu\right)^2}\
\over{2\sigma^2}}\right], \ \ -\infty\ < x < +\infty , \
\end{equation}
where $\mu$ and $\sigma$ are the parameters, called the mean and standard deviation. Fig. \ref{GAUSS_PS} depicts the Gaussian density for $\mu =0$ and $\sigma = 0.5, 1.0$ and $1.5$.  The Gaussian density function plays a central role in the estimation of statistical  errors in Monte Carlo simulation, as we would see later.\\

\noindent
{\bf Exponential distribution:}
\\ 

\begin{figure}[ht]
\centerline{\psfig{figure=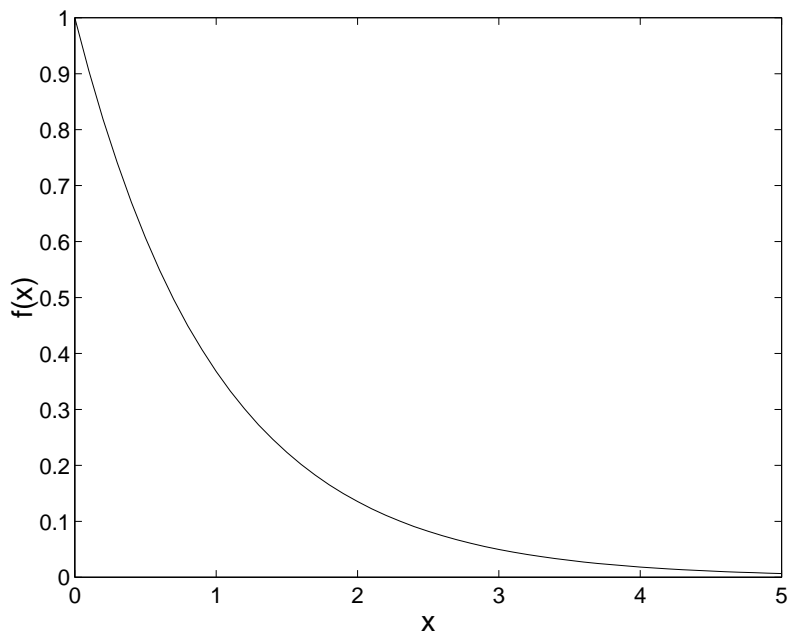,height=8.5cm,width=10.5cm}}
\caption{\protect\small  Exponential distribution, see Eq. (\ref{exp})
with $\alpha =1$.  }
\label{exp_ps}
\end{figure}

\noindent
The {\bf exponential} density arises in the context of several physical phenomena.  The time taken by a radioactive nucleus to decay is exponentially distributed.  The distance a gamma ray photon travels in a medium before absorption or scattering is exponentially distributed. The exponential distribution is given by, \begin{eqnarray}\label{exp}
f(x)=\cases{  \alpha e^{-\alpha x},\ &  for\  $x\ \ge\  0$ \cr &
\cr 0,                     &   for\    $x\  < \     0$ \cr }
\end{eqnarray}
where $\alpha\ >\ 0 $ is a parameter of the exponential distribution. 
Fig. \ref{exp_ps} depicts the exponential distribution, with $\alpha =1$.
\\

\noindent
{\bf Cauchy distribution}
\\

\begin{figure}[ht]
\centerline{\psfig{figure=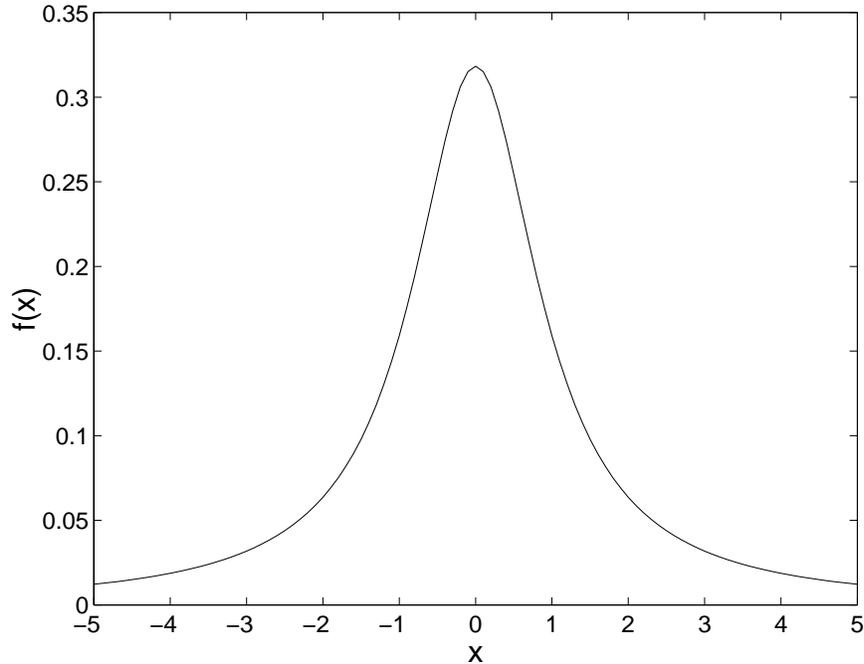,height=9.0cm,width=11.5cm}}
\caption{\protect\small 
Cauchy distribution, see Eq. (\ref{cauchy})  with $D=1$.  }
\label{cau_ps}
\end{figure}

\noindent
An interesting distribution named after Cauchy is given by,
\begin{equation}\label{cauchy}
f(x) ={{1}\over{\pi }} {{D}\over{D^2 + x^2}},\  -\infty\ < x\ <\  +\infty
,
\end{equation} 
where $D > 0$ is a scale factor.  Fig. \ref{cau_ps} 
depicts the {\bf Cauchy} distribution with $D=1$.

If a random variable $\theta$ is uniformly distributed in the range $-\pi/2$ to $+\pi /2$, then $x=\tan (\theta)$ follows a Cauchy distribution ($D=1$). If $X_1$ and $X_2$ are independent Gaussian random variables each with mean zero and standard deviation $\sigma _1$ and $\sigma_2$ respectively, then the ratio $X=X_1 /X_2 $ is Cauchy distributed with $D= \sigma_2 /\sigma_1$.  The Cauchy distribution is also known in the literature as {\bf Lorentz} distribution or  {\bf Breit-Wigner} distribution and  arises in the context of resonance line shapes.

\noindent 
\section{MEAN, VARIANCE AND COVARIANCE}

\noindent
{\bf Mean}
\\

\noindent
The {\bf mean} (also called the expectation value, the average, or the first moment) of a random variable $X(\omega)$, with a  probability density $f(x)$ is denoted by the symbol $\mu$ and is defined as,
\begin{equation}
\mu(X)=\int_{-\infty}^{+\infty}x\  f(x)dx.
\end{equation}
Mean is the most important single parameter in the theory of probability; indeed in Monte Carlo calculations we would be interested in calculating the mean of some random variable, defined explicitly or implicitly. Let us repeat the experiment $N$ times and observe that the outcomes are $\omega_1, \omega_2, \cdots ,\omega_N$. Let the value of the random variable $X(\omega)$ corresponding to these outcomes be $x_1, x_2, \cdots x_N$ respectively. If $N$ is sufficiently large then $(\sum_{k=1}^{N}x_k)/N$  approximately equals $\mu$. In the limit $N\to\infty$ the arithmetic mean of the results of $N$ experiments converges to $\mu$.
\\

\noindent
{\bf Statistical convergence}
\\

\noindent
At this point it is worthwhile discussing the meaning of {\bf statistical convergence} {\it vis-a-vis}  
deterministic convergence that we are familiar with. A given sequence $\{ A(\nu );\ \nu\ =\ 1,\ 2,\cdots\}$ 
is said to converge to a value, say $B$ as $\nu\to\infty$, if for (an arbitrarily small) $\delta > 0$, 
we can find an $M$ such that for all $k\ge M$, $A(k)$ is {\it guaranteed} to be within $\delta$ of $B$ . 
In the statistical context the term {\it guarantee} is replaced by a statement of probability, and the 
corresponding definition of convergence becomes: $A(\nu )$ is said to converge to $B$ as $\nu\to\infty$, 
if given probability $0<P<1$ and (an arbitrarily small) $\delta > 0$, we can find an $M$ such that for all 
$k\ge M$, the probability that $A(k)$ is within $\delta$ of $B$ is greater than $P$. {\it This risk that 
convergence can only be assured with a certain probability is an inherent feature of all Monte Carlo 
calculations.} We shall have more to say on this issue when we take up  discussions on the Chebyshev 
inequality, the law of large numbers and eventually the central limit theorem which help us appreciate 
the nature and content of Monte Carlo errors. For the moment it suffices to say that it is possible 
to make a quantitative probabilistic statement about how close is the arithmetic mean, $A(\nu)$ of $\nu$ 
experiments 
to the actual mean $\mu$, for large $\nu$;  such a statistical error estimate would depend on the number 
of experiments $(\nu )$ 
and the value of the second most important parameter in the theory of probability namely, the 
{\bf variance} of the random variable, $X$, underlying the experiment. 
\\

\noindent
{\bf Variance}
\\ 

\noindent
Variance, $\sigma^2$ is  defined as the expectation value of $(X-\mu)^2$. Formally we have,
\begin{equation}
\sigma^2 (X) = \int_{-\infty}^{+\infty}(x-\mu)^2 f(x) dx .
\end{equation}
The square root of the variance is  called the {\bf standard deviation}.
\\

\noindent
{\bf Moments}
\\

\noindent
We can define, what are called the moments of the random variable. 
The {\bf K}$^{{\rm {\bf th}}}$ {\bf moment} is denoted by $M_K$, and is defined as,
\begin{equation}
M_K = \int_{-\infty}^{+\infty} x^K f(x) dx .
\end{equation}
It is clear that $M_0=1$, which  implies that $f(x)$ is normalized 
to unity, $M_1 =\mu$, and $\sigma^2 = M_2\  -\ M_{1}^2$.
\\

\noindent
{\bf Cumulative probability distribution}\\

\noindent
The {\bf cumulative probability density function}, denoted by $F(x)$ is defined as
\begin{equation}\label{cdf}
F(x) =  {\cal P}\left[ \omega\vert X(\omega)\le x\right] =
\int_{-\infty}^x f(x')dx' .
\end{equation}
$F(x)$ is a monotonic non-decreasing function of $x$; $F(-\infty)=0$ and
$F(\infty)=1$.
\\

\noindent
{\bf Sum of two random variables}\\

\noindent
Let us consider two random variables $X_1$ and $X_2$.  Let $Y_2=X_1 + X_2$.  We have,
\begin{eqnarray}\label{meanoftwo}
\mu(Y_2)&=&\int_{-\infty}^{+\infty}dx_1\int_{-\infty}^{+\infty}dx_2 
(x_1 + x_2 ) f(x_1 , x_2),
\end{eqnarray}
where $f(x_1 , x_2 )$ denotes the {\bf joint density} of the random variables 
$X_1$ and $X_2$. Let $f_1 (x)$ and $f_2 (x)$ denote the probability density functions 
of $X_1$ and $X_2$ respectively. These are called {\bf marginal densities} and are 
obtained from the joint density as follows.
\begin{eqnarray}\label{marginal}
f_1 (x_1) & = & \int_{-\infty}^{+\infty}dx_2 f(x_1 , x_2) .\nonumber\\ f_2
(x_2) & = & \int_{-\infty}^{+\infty}dx_1 f(x_1 , x_2) .
\end{eqnarray}
  The integral in Eq. (\ref{meanoftwo}) can be evaluated and we get,
\begin{eqnarray}
\mu (Y_2) & = & \int_{-\infty}^{+\infty}dx_1\ x_1\int_{-\infty}^{+\infty}dx_2
 \ f(x_1 , x_2) + \nonumber\\ & \  &   \int_{-\infty}^{+\infty}dx_1\
\int_{-\infty}^{+\infty} dx_2\  x_2\  f(x_1,x_2)\nonumber\\ & \    &
\nonumber\\ & = & \int_{-\infty}^{+\infty}dx_1 \ x_1 \ f_1 (x_1) +
\int_{-\infty}^{+\infty}dx_2 \ x_2 f_2 (x_2)\nonumber\\ &  &
\nonumber\\ & =& \mu (X_1) + \mu (X_2).
\end{eqnarray}
The means thus add up. The variance however 
does not, since it involves the square of the random variable.  We have,
\begin{equation}
\sigma^2 (Y_2) = \sigma^2 (X_1) + \sigma^2 (X_2)+ 2\times {\rm cov} [X_1,X_2] .
\end{equation}
The last term in the above is the {\bf covariance} of $X_1$ and $X_2$, and is given by
\begin{equation}\label{covariance}
{\rm cov}(X_1 , X_2)=\int_{-\infty}^{+\infty}dx_1\
\int_{-\infty}^{+\infty}dx_2\ \left[ x_1 -\mu_1\right]\left[x_2 -
\mu_2\right] f(x_1,x_2),
\end{equation}
where $\mu_1 =\mu (X_1 )$ and $\mu_2 =\mu (X_2 )$.  One can define the {\bf conditional 
density} of $X_1$ given that $X_2$ takes a value, say $x_2$, as 
\begin{eqnarray}
f_c (x_1\vert x_2)={{f(x_1,x_2)}\over{f_2 (x_2)}} .
\end{eqnarray}
If $f_c (x_1\vert x_2)=f_1 (x_1)$, then we find that $f(x_1,x_2)=f_1 (x_1)\times f_2(x_2)$.  
The random variables $X_1$ and $X_2$ are then {\bf independent}. In that case we find 
from Eq. (\ref{covariance}) that ${\rm cov}(X_1,X_2)=0$. Thus the covariance is zero if 
the two random variables are independent.  If the covariance  is positive then we say 
that the two random variables are positively correlated; if negative, the random variables 
are  negatively correlated. Note that {\it two random variables may be uncorrelated i.e. 
covariance is zero) but they need not be independent; however if they are independent, 
they must be uncorrelated.} One usually considers the dimension - less normalized quantity 
called the correlation coefficient defined as,
\begin{equation}\label{correcoeff}
C(X_1 ,X_2 )={{  {\rm cov} (X_1, X_2)}\over{ \sigma (X_1) \sigma (X_2) }}
.
\end{equation} 
It is easily verified that $-1 \le C(X_1 , X_2 )\le +1$.
\\

\noindent
Let us calculate the mean and variance of the distributions 
we have seen so far:\\

\newpage
\noindent
{\bf Binomial distribution}\\

\noindent
For the toss of single fair coin,
\begin{eqnarray}\label{binmean}
\mu & = & {{1}\over{2}}\times (+1) + {{1}\over{2}}\times (-1)=0 , \nonumber\\
\sigma^2& =& {{1}\over{2}}\times (+1)^2 + {{1}\over{2}}\times (-1)^2 =1.
\end{eqnarray}

For the toss of $N$ independent  fair coins, the random variable $Y_N =(X_1 + X_2 +\cdots + X_N)$ has 
\begin{equation}
\mu _N={{1}\over{2^N}}
\sum_{n=0}^{N}{{N!}\over{n!(N-n)!}}(2n-N)=0 ,
\end{equation} 
and
\begin{equation}
\sigma^2 _N ={{1}\over{2^N}}\sum_{n=0}^{N}{{N!}\over{n!(N-n)!}}(2n-N)^2=N .
\end{equation}

The random variable $Y_N$ has a simple physical interpretation in terms of a random walk on a 
one-dimensional lattice. The lattice sites are on, say $x$ axis, at unit intervals. You start 
at the origin. Toss a fair coin. If it shows up Heads then step to the right and go to the 
lattice site $x=1$; if toss is Tails, then step to the left and go to the lattice site $x=-1$.  
At the new site, toss the coin again to decide whether to go the left site  or the right site. 
After $N$ tosses find out where you are on the axis, and your position  defines the random 
variable $Y_N$. We can replace the number of coins or the tosses by time $t$ and say $x(t)$ 
is the position of the random walker after time $t$, given it started at origin at time zero. 
Average of $x(t)$ is zero for all times. You do not move on the average. The variance of 
$x(t)$ denoted by $\sigma ^2 (t)$ increases linearly with time $t$,  and the proportionality 
constant is often denoted by $2D$ where $D$ is called the diffusion constant. Thus, for the 
random walk generated by the tossing of fair coins, the diffusion constant is one-half.
 
On the other hand, if we consider the random variable $\bar{Y}_N =Y_N /N$, we find its 
mean is zero and  standard deviation is $1/\sqrt{N}$.  Notice that the standard deviation 
of $\bar{Y}_N$ becomes smaller and smaller as $N$ increases.

Consider now the random variable $\bar{Y}_{\sqrt{N}}=Y_N / \sqrt{N}$; 
it is easily verified that its 
mean is zero and variance is unity, the same as that for the single toss. 
Thus $\sqrt{N}$ seems  to provide a natural scale for  the sum of $N$ independent and 
identically distributed random variables with finite variance. The reason for this would 
become clear in the sequel, see section 8.

For the case of a single loaded coin (with $p$ as the probability for Heads and $q=1-p$ 
as that for the Tails) we have,
\begin{equation}
\mu = p\times (+1) + q\times (-1) = p-q= 2p-1,
\end{equation}
and
\begin{equation}
\sigma ^2 =  p\times (+1)^2 + q\times (-1)^2 - (p-q)^2=4p(1-p).
\end{equation}

For the toss of $N$ independent and identically loaded coins, 
the random variable $Y_N = X_1 + X_2 +\cdots X_N $ has,
\begin{equation}
\mu _N= \sum_{n=0}^{N}{{N!}\over{n!(N-n)!}}p^n q^{N-n}(2n-N)=N( 2p-1),
\end{equation}
and
\begin{equation}
\sigma ^2 _N = \sum_{n=0}^{N}{{N!}\over{n!(N-n)!}}p^n q^{N-n} (2n-N-\mu _N )^2=
4Np(1-p).
\end{equation}
We notice that both the mean and the variance increase linearly with $N$. The relevant quantity 
is the standard deviation relative to the mean; this quantity decreases as $1/\sqrt{N}$. The 
tossing of $N$ loaded coins defines a biased random walk: there is a drift with velocity 
$2p-1$ and a diffusion with  a diffusion constant  $D\equiv \sigma^2/2N =2p(1-p)$.

We can also consider the random variable $\bar{Y}_N = Y_N / N$. Its mean is $2p-1$ 
and is independent of $N$. Its standard deviation, however, decreases with $N$ as $1/\sqrt{N}$. 
Consider now the natural scaling: $\bar{Y}_{\sqrt{N}} =Y_N / \sqrt{N}$; we find its mean is 
$(2p-1)\sqrt{N}$ and variance $4p(1-p)$.  The variance is independent of $N$. \\

\noindent
{\bf Poisson distribution}
\\

\noindent
The mean of the Poisson distribution is given by
\begin{equation}
\mu (t)   =   \sum_{n=0}^{\infty}
n {{ (\lambda t)^n }\over{n!}} e^{-\lambda t} = \lambda t,
\end{equation}
and the variance is given by,
\begin{equation}
\sigma ^2 (t)  =  \sum_{n=0}^{\infty}
(n - \lambda t )^2 {{(\lambda t)^n}\over{n!}}e^{-\lambda t} =  \lambda t .
\end{equation}
Thus, for the Poisson distribution the mean and the variance have the same magnitude.\\
 
\noindent
{\bf Exponential distribution}\\

\noindent 
The mean of the exponential density   is given by
\begin{equation}
\mu = \int_0 ^{\infty}x\alpha\exp(-\alpha x)\ dx\ =\ {{1}\over{\alpha}} ,
\end{equation}
and the variance is,
\begin{equation}
\sigma^2=\int_0 ^{\infty}\left( x-{{1}\over{\alpha}}\right) ^2\alpha
                               \exp(-\alpha x)\ dx= \left(
{{1}\over{\alpha}}\right) ^2 .
\end{equation}

The standard deviation of the exponential density is $\sigma = 1/\alpha$. 
We can associate the standard deviation with something like the expected deviation 
from the mean of a number, sampled randomly from a distribution. We shall 
take up the issue of random sampling from a given distribution later. For the present,
let us assume that we have a large set of numbers sampled independently and randomly 
from the exponential distribution.  The expected fraction of the numbers that fall 
between $\mu - \sigma = 0$ and $\mu +\sigma =2/\alpha$ can be calculated and is given by
\begin{eqnarray}\label{expprob}
{\cal P}(\mu -\sigma \le x \le \mu +\sigma )& =& \int_{\mu-\sigma}^{\mu
+\sigma}
\alpha e^{-\alpha x}\ dx\nonumber\\
                 &   &   \nonumber\\ & = & 0.8647 .
\end{eqnarray}
Thus, nearly  $86\%$ of the sampled numbers are expected to lie within one sigma deviation 
from the mean.  Of course there shall also be numbers much larger than the mean 
since the range of $x$ extends up to infinity.

The distribution of the sum of $N$ independent exponential random variables and the same scaled by $N$ and $\sqrt{N}$ will be considered separately later. In fact I am going to use this example to illustrate the approach to Gaussian as dictated by the Central Limit Theorem.
\\

\noindent
{\bf Cauchy distribution}\\

\noindent
Let us consider the Cauchy distribution discussed earlier. We shall try to evaluate the mean by carrying out the following integration,
\begin{equation}\label{cauchyint}
\mu ={{1}\over{\pi }}
 \int_{-\infty}^{+\infty}x {{D}\over{D^2 + x^2}} \ dx\ .
\end{equation}
The above integral strictly does not exist. The reason is 
as follows. If we carry out the integration from $-\infty$ 
to $0$ and then from $0$ to $\infty$, these two integrals 
do not exist. Notice that the integrand is a odd function. Hence 
if we allow something like 
a {\it principal} value integration, where the limits are 
taken simultaneously, we can see that the integral 
from $-\infty$ to $0$ cancels out that from $0$ to 
$+\infty$, and we can say the mean, $\mu $  is zero, 
consistent with the graphical interpretation of the density, see 
Fig. \ref{cau_ps}.  If we now try to evaluate the variance, we find,
\begin{equation}\label{cauvar}
\sigma ^2 = {{1}\over{\pi}}
 \int_{-\infty}^{+\infty}(x-\mu)^2 {{D}\over{D^2 + x^2}}\  dx.
\end{equation}
The above is an unbounded integral. Hence if we sample numbers randomly 
and independently from the Cauchy distribution and make an attempt 
to predict the extent to which these  numbers fall close 
to the \lq mean\rq\ , we would fail. Nevertheless
Cauchy distribution is a legitimate probability distribution since its integral from $-\infty$ to $+\infty$ is unity and for all values of $x$, the distribution function is greater than or equal to zero.  But its variance is infinity and its mean calls for a \ \lq generous\rq\ interpretation of the integration.  Since the standard deviation for the Cauchy distribution is infinity, the width of the distribution is usually characterized by the {\bf Full Width at Half Maximum(FWHM)} and it is $2D$.

Consider the sum $Y_N = X_1 + X_2 + \cdots + X_N$ of $N$ independent Cauchy random variables. What is the probability distribution of the random variable $Y_N$~? Does it tend to Gaussian as $N\to\infty $~?  If not, why~?  What is the natural scaling for the sum $Y_N$~?  These and related questions we shall take up later. Bear in mind that the Cauchy distribution has unbounded variance and this property will set it apart from the others which have  finite variance.
\\

\noindent
{\bf Gaussian distribution}\\

\noindent
For the Gaussian distribution the two parameters $\mu$ and $\sigma$ are the mean and standard deviation respectively. Let us calculate the probability that a number sampled from a Gaussian falls within one sigma interval around the mean. This can be obtained by integrating the Gaussian from $\mu -\sigma$ to $\mu +\sigma$. We get,
\begin{eqnarray}\label{gaussint}
{\cal P}(\mu -\sigma \le x\le \mu +\sigma )&=&
{{1}\over{\sigma\sqrt{2\pi}}}\int_{\mu -\sigma }^{\mu +\sigma }\exp
\left[- {{ \left( x-\mu\right) ^2}\over{2\sigma ^2}}\right] \ dx ,
\nonumber\\
                                          &   & \nonumber\\ & = & 0.6826.
\end{eqnarray}
Thus $68\%$ of the numbers sampled independently and randomly from a Gaussian distribution  are expected to fall within  one sigma interval around the mean. This interval is usually called the {\it one-sigma confidence} interval.

The sum $Y_N$ of $N$ independent Gaussian random variables is also a Gaussian with mean $N\mu$ and variance $N\sigma^2$. When you scale the sum by $N$, the mean is $\mu$ and the variance is $\sigma^2 /N$; on the other hand the random variable $Y_N /\sqrt{N}$, has mean $\mu\sqrt{N}$ and variance $\sigma^2$. These will become clear when we take up the discussion on the Central Limit theorem later.  In fact  we  shall see that the arithmetic mean of $N$ independent random variables, each with finite variance, would tend to have a Gaussian distribution for large $N$. The standard deviation of the limiting Gaussian distribution will be proportional to the inverse of the square root of $N$.  We shall interpret the one-sigma confidence interval of the Gaussian as the statistical error associated with  the estimated mean of the $N$ independent random variables.
\\  

\noindent
{\bf Uniform distribution}\\

\noindent
For the uniform random variable $U(a,b)$, the mean is,
\begin{equation}\label{meanuni}
\mu = {{1}\over{b-a}}\int_a ^b x\ \ dx = {{a+b}\over{2}},
\end{equation}
and the variance is,
\begin{equation}\label{varuni}
\sigma ^2 ={{1}\over{b-a}} \int_a ^b \left( x-{{a+b}\over{2}}\right)^2 \ dx = {{(b-a)^2}\over{12}}.
\end{equation}
 For the random variable $U(0,1)$, the mean and variance are respectively
$1/2$ and $1/12$.

\begin{boxedtext}
\noindent
{\bf Assignment 3}\\

\noindent
Consider $N$ independent uniform random variables $U(0,1)$.  Let $Y_N$
denote their sum.\\
\noindent
(A) Find the mean and variance of (a) $Y_N$, (b) $Y_N /N$ and (c) $Y_N
/\sqrt{N}$ as a function $N$.\\
\noindent
(B) Derive an expression for the probability distribution of $Y_N$ for
$N=2$.
\end{boxedtext}

\noindent
\section{CHARACTERISTIC FUNCTION}

\noindent
The  {\bf characteristic function} of a random variable $X$ is defined as
the expectation value of $\exp(ikX)$.
\begin{equation}
\Phi_X (k) = \int_{-\infty}^{+\infty}\exp (ikx) f(x)\ dx .
\end{equation}
Expanding the exponential we get,
\begin{equation}
\Phi_X (k)=\int_{-\infty}^{+\infty}f(x)\left[1+(ikx)+{{(ikx)^2}\over{2!}}+
        \cdots + {{(ikx)^n }\over{n!}}+\cdots\right]\ dx .
\end{equation}
Assuming that term wise integration is valid, we find,
\begin{equation}\label{MGF_EQ}
\Phi_X (k) = 1+ikM_1+{{(ik)^2}\over{2!}}M_2+\cdots+{{(ik)^n}\over{n!}}M_n
\cdots  , 
\end{equation}
from which we get,
\begin{equation}
i^n M_n =\left. {{d^n}\over{dk^n }} \Phi_X (k)\right\vert _{k=0}  .
\end{equation}
Thus we can generate all the moments from the characteristic function.

For a discrete random variable, the characteristic function is defined
similarly as,
\begin{equation}
\Phi (k) = \sum_{n} \exp (ikx_n)p_n .
\end{equation}
The logarithm of the characteristic function generates, what are called the {\bf cumulants} or the {\bf semi-invariants}. We have,
\begin{equation}\label{CGF_EQ}
\Psi_X (k) = \ln\left[  \Phi_X (k)\right] \equiv \sum_{n=1}^{\infty} 
            {{ (ik)^n}\over{n!}} C_n,
\end{equation}
where $C_n$ is the $n$-th cumulant, given by,
\begin{equation}
i^n C_n =\left.{{d^n}\over{dk^n}}  \Psi_X (k)\right\vert_{k=0} .
\end{equation}\\

\noindent
{\bf  How are the cumulants related to the moments~?}\\

\noindent
This can be found by considering,
\begin{eqnarray}
\ln \left[ 1+ikM_1 + {{(ik)^2}\over{2!}}M_2 + {{ (ik)^3 }\over{3!}}M_3+
\cdots\right]
=\quad\quad\quad\quad\quad\nonumber\\ ikC_1 + {{(ik)^2}\over{2!}} C_2 +
{{(ik)^3}\over{3!}} C_3 + \cdots .
\quad\quad\quad\quad\quad\quad\quad\quad
\end{eqnarray}
Taylor-expanding the logarithm on the left hand side of the above and equating the coefficients of the same powers of $k$, we get,
\begin{eqnarray}
C_1 & = & M_1\\ C_2 & = & M_2 - M_1 ^2 = \sigma ^2\\ C_3 & =& M_3 - 3M_2
M_1 +2 M_1 ^3 ,\\ C_4 & = & M_4 - 4M_3 M_1 - 3M_2 ^2 + 12M_2 M_1 ^2 - 6
M_1 ^4  .
\end{eqnarray}
The first cumulant is the same as the first moment (mean). The second
cumulant is the variance.

A general and simple expression relating the moments to the cumulants can
be obtained \cite{VALSA} as follows. We have,
\begin{equation}
{{d\Phi(k)}\over{dk}}=\Phi(k){{d\Psi(k)}\over{dk}},
\end{equation}
where $\Phi(k)$, see Eq. (\ref{MGF_EQ}) and $\Psi(k)$, see Eq.
(\ref{CGF_EQ}) are the moments and cumulants generating functions
respectively; we have dropped the suffix $X$ for convenience. We see that,
\begin{eqnarray}
{{d^{n+1}\Phi(k)}\over{dk^{n+1} }}&=&{{d^n}\over{dk^n}}\left[ \Phi(k)
{{d\Psi(k)}\over{dk}}\right]\nonumber\\ & = & \sum_{m=0}^{n}
{{n!}\over{m!\ (n-m)!}} {{d^{n-m} \Phi(k)}\over{dk^{n-m}}} {{d^{m+1}\Psi
(k)}\over{dk^{m+1}}},
\end{eqnarray}
from which it follows that,
\begin{eqnarray}
C_1 & = & M_1 \nonumber\\ C_{n+1}&=&M_{n+1} - \sum_{m=0}^{n-1}
{{n!}\over{m!\ (n-m)!}} M_{n-m}C_{m+1}
\ \ \forall\ n\ >\ 0.
\end{eqnarray}
\\

\noindent
Let us calculate the characteristic functions of the 
several random variables considered so far:\\

\newpage
\noindent
{\bf Binomial distribution}\\

\noindent
The characteristic function of the random variable defined for  the toss of a single fair coin is
\begin{eqnarray}
\Phi_X (k)& = & {{1}\over{2}} e^{+ik}+{{1}\over{2}} e^{-ik}\nonumber\\
\         & = & \cos (k).
\end{eqnarray}
The characteristic function of $Y_N = X_1 + X_2 +\cdots X_N$, defined for the toss of $N$ independent fair coins is given by,
\begin{eqnarray}
\Phi_{Y_N} (k)& = & {{1}\over{2^N }} \sum_{n=0}^{N}  {{N!}\over{n!(N-n)!}}
           e^{ik(2n-N)}\nonumber\\ & & \nonumber\\
&=&{{1}\over{2^N}}\sum_{n=0}^{N} {{N!}\over{n! (N-n)!}} \left(
e^{ik}\right) ^n \left( e^{-ik}\right) ^{N-n}\nonumber\\ & & \nonumber\\
&=&\left( {{e^{ik} + e^{-ik} }\over{2}}\right) ^N \nonumber\\ &
&\nonumber\\
\      &=& \left[ \cos (k)\right] ^N \nonumber\\
\      &=& \left[ \Phi_X (k)\right]^N .
\end{eqnarray}

\noindent
{\bf Sum of independent random variables}\\

\noindent
A general result is that the characteristic function of the sum of independent random variables is given by the product of the characteristic functions of the random variables. Let $X_1 , X_2 , \cdots X_N$ be independent and not necessarily 
identically distributed.  Let $Y_N=X_1 + X_2 + \cdots +X_N$. Formally, the characteristic function of $Y_N$, denoted by $\Phi_{Y_N} (k)$ is given by
\begin{eqnarray}\label{phiy}
\Phi_{Y_N} (k)=\int dx_1\  \int dx_2\  \cdots\quad\quad\quad\quad
\quad\quad\quad\quad\quad\quad\quad\nonumber\\
         \cdots\int dx_N\
\exp \left[ ik\left( x_1 +x_2 + \cdots +x_N \right) \right] f(x_1,x_2,\cdots
,x_N).
\end{eqnarray}
Since $X_1, X_2, \cdots , X_N $ are independent, the joint density can be
written as the product of the densities of the random variables. {\it
i.e.},
\begin{equation}
f(x_1 ,x_2 ,\cdots x_N )=f_1 (x_1 )f_2 (x_2 )\cdots f_N (x_N ) ,
\end{equation}
where $f_i(x_i)$ is the probability density of $X_i$. Therefore,
\begin{eqnarray}
\Phi_{Y_N}(k)&=&\prod_{i=1}^{N}\int dx_i\  
      \exp\left( ikx_i \right) f_i(x_i )\nonumber\\ & &
\nonumber\\
\        &=&\prod_{i=1}^{N}\Phi_{X_{i}} (k ),
\end{eqnarray}  
where $\Phi_{X_{i}}(k)$ denotes the characteristic function of the random variable $X_i$. If these  $N$ random variables are also identical, then $\Phi_{Y_N} (k)= [\Phi_X (k)]^N$.
\\

\noindent
{\bf Arithmetic mean of independent and identically 
distributed random variables}
\\ 

\noindent
Let
\begin{equation}
\bar{Y}_N = {{ X_1 + X_2+ \cdots  + X_N}\over{N}},
\end{equation}
define the arithmetic mean of  $N$ independent and 
identically distributed random variables. In Eq. (\ref{phiy}), replace $k$ by $k/N$; the resulting equation defines the characteristic function of $\bar{Y}_N$.  Thus we have,
\begin{equation}
\Phi_{\bar{Y}_N} (k)= \Phi_{Y_N} (k/N) = \left[ \Phi_X (k/N)\right]^N.
\end{equation}

\noindent
{\bf Sum of $N$ independent and identically distributed 
random variables scaled by $\sqrt{N}$}\\
 
\noindent
Later, we shall have occasions to consider the random variable $\bar{Y}_{\sqrt{N}}$, defined by,
\begin{equation}
\bar{Y}_{\sqrt{N} }={{ X_1 + X_2 + \cdots X_N}\over{\sqrt{N} }},
\end{equation}
whose characteristic function can be obtained by replacing $k$ by
$k/\sqrt{N}$ in Eq. (\ref{phiy}). Thus,
\begin{equation}
\Phi_{\bar{Y}_{\sqrt{N} } }(k) = \Phi_{Y_N}(k/\sqrt{N})=
\left[ \Phi_X (k/\sqrt{N})\right] ^N .
\end{equation}
\\

\noindent
{\bf Exponential distribution}\\

\noindent
For the exponential distribution, see Eq. (\ref{exp}), with mean unity ({\it i.e.}\ $ \alpha~=~1$), the characteristic function is given by,
\begin{eqnarray}
\Phi_X (k) & = & \int_{0}^{\infty} e^{ikx-x}\ dx\nonumber\\
\          & = & {{1}\over{1-ik}}\ \  .
\end{eqnarray}
The sum $Y_N$ of $N$ exponential random  variables has a characteristic function given by,
\begin{equation}\label{sumexp}
\Phi_{Y_N} (k) = {{1}\over{ (1-ik)^N}}\quad .
\end{equation}
The random variable $\bar{Y}_N$ has a characteristic function given by, \begin{equation}\label{amexp}
\Phi_{\bar{Y}_N}(k) = {{1} \over{ \left( 1- {{ik}\over{N}} \right)^N}}\quad ,
\end{equation}
which has been obtained by replacing $k$ by $k/N$ in Eq. (\ref{sumexp}).
\\

\noindent
{\bf Poisson distribution}
\\

\noindent
For the Poisson distribution, see Eq. (\ref{poissondist}), the characteristic function is given by,
\begin{equation}\label{poissonch}
\Phi (k,t)=\sum_{n=0}^{\infty} {{e^{ikn} (\lambda t)^n e^{-\lambda t}}\over{n!}}
          = \exp\left[ -\lambda t \left( 1-e^{ik}\right)\right],
\end{equation}
which is the same as Eq. (\ref{poigen}) if  we set  $z=\exp (ik)$.\\

\noindent
{\bf Gaussian distribution}\\

\noindent
For the Gaussian distribution, the characteristic function is given by,
\begin{equation}\label{ftgauss}
\Phi_X (k) = \exp(i\mu k-{{1}\over{2}}\sigma^2 k^2) .
\end{equation}
Following the rules described above, we have,
\begin{eqnarray}
\Phi_{Y_N} (k) & =& \exp \left[ iN\mu k - {{1}\over{2}}N\sigma^2 k^2\right]
\label{sumgauss1}\\
\Phi_{\bar{Y}_N} (k) & = &  \exp\left[ i\mu k -
         {{1}\over{2}}{{\sigma^2 k^2}\over{N}}\right]
\label{sumgauss2}\\
\Phi_{\bar{Y}_{\sqrt{N}}} (k) & = &  \exp\left[ i\sqrt{N}\mu k -
         {{1}\over{2}}\sigma^2 k^2\right] .
\label{sumgauss3}
\end{eqnarray}
For a Gaussian, only the first two cumulants are non-zero. 
identically zero.  From Eq. (\ref{sumgauss3}) we see that the sum of $N$ Gaussian random variables scaled by $\sqrt{N}$, is again a Gaussian with variance independent of $N$.

\begin{boxedtext}
\noindent
{\bf Assignment 4}\\

\noindent
Derive expressions for the characteristic functions of (a)~the Gaussian and b)~the Cauchy distributions.
\end{boxedtext}
\noindent
\section{CHEBYSHEV INEQUALITY}

\noindent
Let $X$ be an arbitrary random variable with a probability density function $f(x)$, and finite variance $\sigma^2$. The {\bf Chebyshev inequality}, see Papoulis \cite{PAP}, says that, 
\begin{equation}
{\cal P}\left\{ \vert X-\mu\vert \ge k\sigma \right\} \le {{1}\over{k^2}}\
\ ,
\end{equation}
for $k\ge 1$.  Thus, regardless of the nature of the density function $f(x)$, the probability that the random variable $X$ takes a value between $\mu-\epsilon$ and $\mu+\epsilon$, is greater than $1-(\sigma^2 / \epsilon^2)$, for $\epsilon\ge\sigma$. The Chebyshev inequality follows directly from the definition of the variance, as shown below.
\begin{eqnarray}
\sigma^2 & =    & \int_{-\infty}^{+\infty}(x-\mu)^2 f(x)\ dx\nonumber\\
\        & \ge  & \int_{-\infty}^{\mu-k\sigma}(x-\mu)^2 f(x)\ dx +
                   \int_{\mu +k\sigma}^{+\infty}(x-\mu)^2 f(x)\
dx\nonumber\\
\        & \ge  & k^2 \sigma^2\left[ \int_{-\infty}^{\mu-k\sigma}f(x)\ dx +
                   \int_{\mu+k\sigma}^{+\infty}f(x)\ dx\
\right]\nonumber\\
\        &  =   & k^2\sigma^2{\cal P}
                  \left\{ \left\vert X-\mu\right\vert \ge k\sigma\right\}\
\  .
\end{eqnarray} 
The Chebyshev inequality is simple; it is  easily adapted to sums of random variables and this precisely is of concern to us in the Monte Carlo simulation. Take for example, $N$ independent realizations of a random variable $X$ with mean zero and variance $\sigma^2$. Let $\bar{Y}_N$ be the arithmetic mean of these realizations. $\bar{Y}_N$ is a random variable. The mean of $\bar{Y}_N$ is zero and its variance is $\sigma^2 /N$. Chebyshev inequality can now be applied to the random variable $\bar{Y}_N$. Accordingly, a particular realization of the random variable $\bar{Y}_N$ will lie outside the interval $(-\epsilon , +\epsilon )$ with a probability less than   or equal to $\sigma^2/(N\epsilon^2)$.  Thus, as $\epsilon$ becomes smaller, by choosing $N$ adequately large we find that a realization of $\bar{Y}_N$ can be made to be as close to the mean as we desire with a probability very close to unity.

This leads us naturally to the {\bf laws of large numbers}. Of the several laws of large numbers, discovered  over a period two hundred years, we shall see perhaps the earliest version, see Papoulis \cite{PAP}, which is, in a sense, already contained in the Chebyshev inequality.

\noindent
\section{LAW OF LARGE NUMBERS}

\noindent
Consider the  random variables $X_1 , X_2 , \cdots , X_N$ which are independent and 
identically distributed. The  common probability density has a mean $\mu$ and a {\it finite} variance. Let $\bar{Y}_N$ denote the sum of the random variables divided by $N$. The law of large numbers says that for a given $\epsilon > 0$, as $N\to\infty$,
\begin{eqnarray}
{\cal P}  \left\{ \left\vert\bar{Y}_N -\mu\right\vert\ > \epsilon\right\}
& \to 0 &  .\nonumber
\end{eqnarray}
It is easy to see that a realization of the random variable $\bar{Y}_N$ is just the Monte Carlo estimate of the mean $\mu$ from a sample of size $N$. The law of large numbers assures us that the {\bf sample mean} converges to the {\bf population mean} as the sample size increases. I  must emphasize that the convergence we are talking about here is in a probabilistic sense. Also notice that the law of large numbers does not make any statement about the nature of the probability density of the random variable $\bar{Y}_N$. It simply assures us that in the limit of $N\to\infty$, the sample mean would converge to the right answer. $\bar{Y}_N$ is called the {\bf  consistent estimator} of $\mu$.

 The central limit theorem on the other hand, goes a step further and tells us about the nature of the probability density function of $\bar{Y}_N $, as we shall see below.

\noindent
\section{CENTRAL LIMIT THEOREM (CLT)}

\noindent
Let $X_1 , X_2 , \cdots X_N$ be $N$ independent and 
identically distributed random variables having a common Gaussian probability density with mean zero and variance $\sigma^2$, given by,
\begin{equation}
G_{X} (x)={{1}\over{\sigma\sqrt{2\pi}}}\exp \left[ - {{x^2}\over{2\sigma
^2}}\right].
\end{equation}
Let $Y_N =X_1 + X_2 + \cdots X_N $. It is clear that the characteristic function of $Y_N$ is the characteristic function of $X$ raised to the power $N$. Thus, from Eq. (\ref{ftgauss}),
\begin{eqnarray}
\Phi_{Y_N} (k) & \equiv & \left[ \Phi_X (k)\right] ^N \nonumber\\
\          &   =    & \exp \left[ -{{1}\over{2}}\sigma ^2
                                          k^2 N\right] .
\end{eqnarray}
Fourier inverting the above, we find that the probability density of the random variable $Y_N$ is Gaussian given by,
\begin{equation}
G_{Y_N} (x)={{1}\over{\sigma\sqrt{2\pi N} }}\exp\left[- 
                     {{x^2 }\over{2N\sigma^2}}\right] ,
\end{equation}
with mean zero and variance $N\sigma^2$.  Thus when you add Gaussian random variables, the sum is again a Gaussian random variable. Under addition, Gaussian is stable. The scaling behaviour of the distribution is evident,
\begin{equation}
G_{Y_N} (x) = N^{-1/2}G_X \left( {{x}\over{N^{1/2} }}\right) \ .
\end{equation}

The variance of the sum of $N$ independent and 
identically distributed Gaussian random variables increases (only) linearly with $N$. On the other hand the variance of $\bar{Y}_N$ is  $\sigma^2 /N$, and thus decreases with $N$.  Therefore  the probability density of $\bar{Y}_N$ is,
\begin{equation}\label{clt}
G_{\bar{Y}_{\sqrt{N}}} ={{\sqrt{N} }\over{\sigma\sqrt{2\pi} }}\exp
             \left[ - {{Nx^2}\over{2\sigma^2}}\right] .            
\end{equation}
{\bf The Central Limit Theorem asserts that even if the common probability density of the $N$ random variables is not Gaussian, but some other arbitrary density with (zero mean and) finite variance,  Eq. (\ref{clt}) is still valid but  in the limit of $N\to\infty$.} To see this, write the characteristic function of the common probability density as
\begin{eqnarray}
\Phi_X (k) & = & \int_{-\infty}^{+\infty}e^{ikx} f(x)\ dx\nonumber\\
\          & = & 1-{{1}\over{2}}\sigma^2 k^2 - {{1}\over{6}}iM_3 k^3
                +{{1}\over{24}}M_4 k^4  \cdots
\end{eqnarray}
Hence the characteristic function of $\bar{Y}_N$ is,
\begin{eqnarray}
\Phi_{\bar{Y}_N} (k)&=& \left[ \Phi_X (k/N)\right] ^N\nonumber\\
\         &\approx& \left[ 1-{{1}\over{2}} {{\sigma^2 k^2}\over{N^2}}
               \left( 1+i{{M_3 k}\over{3\sigma^2}} {{1}\over{N}}
               -{{M_4 k^2}\over{12\sigma^2}}{{1}\over{N^2}}\cdots  \right)
                \right]^N\nonumber\\
\         &\sim  & \exp\left[- {{1}\over{2}} {{\sigma^2 k^2}\over{N}}\right]
\quad (\rm{for}\quad N\to\infty) , 
\end{eqnarray}
whose inverse is the density given by Eq. (\ref{clt}), see van Kampen
\cite{VANK}.

The above can be expressed in terms of cumulant expansion. We have for the random variable $X$,
\begin{eqnarray}
\Phi_X (k)&=&\exp [ \Psi_X (k)]\nonumber\\
 &=& \exp\left[ ikC_1 + {{(ik)^2}\over{2!}} C_2 + 
           {{ (ik)^3}\over{3!}}C_3 
+\cdots\right] ,
\end{eqnarray}
where $C_1 , C_2 , C_3\cdots $ are the cumulants of $X$.  Then, the characteristic function of the random variable $\bar{Y}_N$ is given by,
\begin{eqnarray}
\Phi_{\bar{Y}_N} (k)=\left[ \Phi_X  (k/N)\right]^N\quad\hskip 6 cm\nonumber\\
                     =\exp\left[ ikC_1 - 
            {{1}\over{2}} {{k^2 C_2}\over{N}}\left(
1+{{ik}\over{3}}{{C_3 }\over{C_2}}{{1}\over{N}}
-{{k^2 C_4}\over{12C_2}}{{1}\over{N^2}}     \cdots \right) \right] .
\end{eqnarray}
We immediately see that for $N\to\infty$,
\begin{equation}
\Phi_{\bar{Y}_N} \sim  \exp\left[ i\ k\ C_1 - {{1}\over{2}} {{k^2 C_2}\over{N}}\right],
\end{equation}
whose Fourier-inverse is a Gaussian with mean $C_1$, and variance $C_2 /N$.

For the random variable $\bar{Y}_{\sqrt{N} }$, defined as the sum of $N$ random variables divided by $\sqrt{N}$, the characteristic function can be written as,
\begin{eqnarray}
\Phi_{\bar{Y}_{\sqrt{N} }} = \left[ \Phi_X (k/\sqrt{N})\right]^N\hskip  65 mm \nonumber\\
                         = \exp\left[ ik\sqrt{N}C_1 -
                                          {{k^2}\over{2}} C_2
                 \left\{ 1 + {{ ikC_3 }\over{3C_2}}{{1}\over{ \sqrt{N} }}
               - {{k^2 C_4}\over{12C_2}}{{1}\over{N}}
                                             +\cdots\right\}\right]\nonumber\\
          \sim \exp\left[ ik\sqrt{N}C_1-
     {{k^2 C_2}\over{2}}\right] \quad (\rm{ for}\quad N\to\infty),\hskip 27 mm 
\end{eqnarray}
which upon Fourier inversion gives a Gaussian with mean $C_1\sqrt{N}$ and variance $C_2$. The variance of $\bar{Y}_{\sqrt{N}} $ is independent of $N$.

\noindent
\subsection{L\'evy Distributions}

\noindent
In the beginning of the discussion on the  Central Limit Theorem, I said when you add Gaussian random variables, what you get is again a  Gaussian. Thus Gaussian  is {\it stable} under addition.  This is a remarkable property. Not all distributions have this property. For example, when you add two uniform random variables, you get a random variable with triangular distribution.  When two exponential random variables  are added we get one with  a Gamma distribution, as we would see in the next section where we shall be investigating the behaviour of the sum of several independently distributed exponential random variables.  (In fact we went a step further and found that when we  add $N$ 
identically distributed independent random variables with finite variance, the resulting distribution  tends to a Gaussian, when $N\to\infty$. This is what we called the Central Limit Theorem).

A natural question that arises in this context is: Are there any other distributions, 
that are stable under addition~? The answer is yes and they are called the {\bf L\'evy stable distributions}, discovered by Paul L\'evy \cite{levy} in the mid twenties.  Unfortunately the general form of the L\'evy distribution is not available. What is available is its characteristic function. Restricting ourselves to the symmetric case, the characteristic function of a L\'evy distribution is given by,
\begin{equation}
{\tilde L}(k;\alpha )=\exp (-D\vert k\vert ^{\alpha} ),
\end{equation}
where $D > 0 $ is a scale factor, $k$ is the transform variable corresponding to $x$,  and $\alpha $ is the L\'evy index.  The Fourier inverse of ${\tilde L}(k;\alpha )$ is the L\'evy distribution $L(x;\alpha )$.  L\'evy showed that for $L(x;\alpha )$ to be non negative, $0<\alpha\le2$.  In fact we have,
\begin{eqnarray}
L(x;\alpha )\sim\cases{
                   D^{-1/\alpha}\ ,\ \  &  for \  $x=0$;\cr
     D\vert x\vert^{-\alpha-1}\ ,\  & for \  $\vert x\vert \to\infty ;\  
\alpha < 2$ \ ;\cr
    D^{-1/2}\exp (-x^2 / 4D) \ &$ ,\forall\  x$   and \   $\alpha =2$\cr}.
\end{eqnarray}
The pre factor in the above can be fixed by normalization. We see that Gaussian is a special case of the  L\'evy distributions when $\alpha =2$. In fact Gaussian is the only L\'evy distribution with finite variance. Earlier we saw about Cauchy distribution. This is again a special case of the L\'evy distribution obtained when we set $\alpha =1$.

Consider $N$ independent and identically distributed L\'evy random variables, with the common distribution denoted by $L(x;\alpha)$. The sum $Y_N$ is again a L\'evy distribution denoted by $L_N (x; \alpha)$, obtained from $L(x;\alpha)$ by replacing $D$ by $ND$. Let us scale $Y_N$ by $N^{1/\alpha}$ and consider the random variable $Y_N / N^{1/\alpha}$. Its  distribution is $L(x; \alpha)$. Thus $N^{1/\alpha}$ provides a natural scaling for Levy distributions. We have,
\begin{equation}
L_{N}  (x; \alpha)=N^{- 1/ \alpha } L_X  \left( {{x}\over{ N^{1/\alpha}} } ; \alpha\right) .
\end{equation}
The scaling behaviour fits into a general scheme. The Gaussian corresponds to $\alpha =2$; the natural scaling for Gaussian is thus $\sqrt{N}$. The Cauchy distribution corresponds to $\alpha =1$; the natural scaling is given by $N$.

Thus L\'evy stable law generalizes the central limit theorem: L\'evy distributions are the only possible limiting distributions for the sums of independent identically distributed random variables. The conventional Central Limit Theorem is a special case restricting each random variable to have finite variance and we get Gaussian as a limiting distribution of the sum. L\'evy's more general Central Limit Theorem applies to sums of random variables with finite or infinite variance.

Stable distributions arise in a natural way when 1) a physical system evolves, the evolution being  influenced by additive random factors and 2) when the result of an experiment is determined by the sum of a large number of independent random variables.

\noindent 
\subsection{Illustration of the  CLT} 

\noindent
I  shall illustrate the approach to Gaussian of the distribution of the mean, by considering exponential random variables for which all the relevant quantities can be obtained analytically. We start with the identity,
\begin{equation}
\int_{0}^{\infty}\ dx\  e^{-\alpha x}\ e^{ikx} = {{1}\over{\alpha -ik}}.
\end{equation}  
We  differentiate $(N-1)$ times, both sides of the above  with respect to $\alpha$ and set $\alpha =1$. We get,
\begin{equation}\label{gfsum}
\int_{0}^{\infty}dx\  e^{ikx}\ \left[ {{e^{-x}x^{N-1}}\over{(N-1)!}}\right]
={{1}\over{(1-ik)^N}}.
\end{equation}
We immediately identify the right hand side of the above, as the characteristic function of the sum of $N$ independent and identically distributed exponential random variables with mean unity, see Eq. (\ref{sumexp}). In Eq. (\ref{gfsum}) above, we replace, on both sides, $k$ by $k/N$, $x$ by $Nx$, and $dx$ by $Ndx$, to get,
\begin{equation}\label{phiam}
\int_0 ^{\infty} dx\  e^{ikx}\left[ {{ e^{-Nx}N^N x^{N-1} }\over{(N-1)!}}
\right] = {{1}\over{\left( 1- i{{k}\over{N}}\right)^N }}.
\end{equation}
The right hand side of the above is the characteristic function of $\bar{Y}_N$  see Eq. (\ref{amexp}).  Thus we get an exact analytical expression for the probability density function of $\bar{Y}_N$, as
\begin{equation}
f(x)={{N^N}\over{(N-1)!}}\ e^{-Nx}\ x^{N-1}, \quad\quad 0\le x < \infty , 
\end{equation}
which, for all $N$,  is a {\bf gamma density function}, and not a Gaussian!

The cumulants of the gamma density can be obtained by power series expansion of the logarithm of the characteristic function  given by Eq. (\ref{amexp}). We get,
\begin{eqnarray}
\ln \Phi_{\bar{Y}_N}(k)&=&-N\ln \left( 1- {{ik}\over{N}}\right)\nonumber\\
                    &=&N\sum_{n=1}^{\infty} {{ (ik)^n }\over{ N^n n}}\nonumber\\
                    &                                                \nonumber\\
                    &=&\sum_{n=1}^{\infty} {{(ik)^n}\over{n!}}
                       \left[ {{ (n-1)!}\over{N^{n-1} }}\right],\nonumber\\
                    &                                           \nonumber\\
                   & =&ik+ {{ (ik)^2}\over{2!}} {{1}\over{N}}
                           \left[ 1+ {{ ik}\over{3}}{{2!}\over{N}
                           }\cdots \right]  .
\end{eqnarray}
The $n$-th cumulant is thus,
\begin{equation}
C_n = {{ (n-1)!}\over{N^{n-1} }}.
\end{equation}
We find $C_1 =1$ and $C_2 = 1/N$, as expected. The third cumulant  is $C_3 = 2/N^2$, and it goes to zero as $N\to\infty$. In fact all the higher cumulants vanish in the asymptotic limit. (Indeed even the second cumulant goes to zero in the limit $N\to\infty$, which means that the probability density of the arithmetic mean is asymptotically a delta function, something very satisfying from the point of view of a Monte Carlo practitioner!).  The important  point is that the Gaussian with mean $unity$ and variance $1/N$ becomes a very good approximation to the gamma density, for large $N$.

\begin{figure}[ht]
\centerline{\psfig{figure=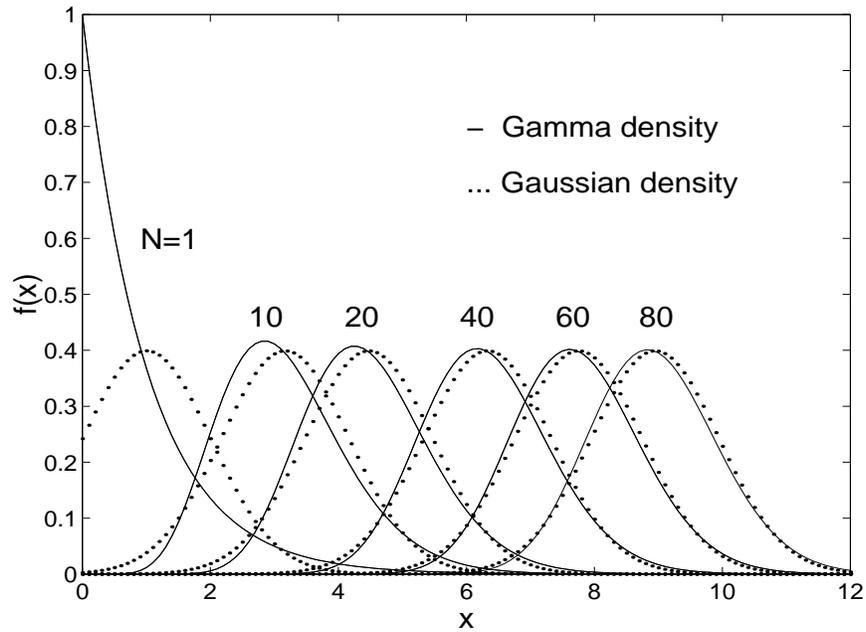,height=8.5cm,width=11.5cm}}
\caption{\protect\small Gamma density  and the asymptotic
Gaussian}
\label{GAMMA_PS}
\end{figure}

Replacing $k$ by $k/\sqrt{N}$, $x$ by $x\sqrt{N}$ and  $dx$ by $dx\sqrt{N}$ in Eq. (\ref{gfsum}), we get, 
\begin{equation}
\int_0 ^{\infty} dx\  e^{ikx}\left[ {{(\sqrt{N})^N}\over{(N-1)!}}\ x^{N-1}
\  e^{ -\sqrt{N}x }\right] =
{{1}\over{ \left( 1- i {{k}\over{\sqrt{N} }} \right) ^N}} .
\end{equation}
The right hand side of the above equation is the characteristic function of the random variable $\bar{Y}_{\sqrt{N}}$ $=$ $ (X_1 + X_2 + \cdot + X_N )/\sqrt{N}$.  Thus, the probability density function of $\bar{Y}_{\sqrt{N} }$, is given by the gamma density,
\begin{equation}\label{gammarootn}
f(x) = {{ (\sqrt{N})^N}\over{(N-1)!}}\ x^{N-1}
\  \exp (-\sqrt{N}x ),
\end{equation}
whose mean is $\sqrt{N}$ and whose variance is unity (independent of $N$). 
Fig. \ref{GAMMA_PS} depicts the gamma density given above for 
$N=1$, $10$, $20$, $40$, $60$ and $80$ along with the Gaussian of
mean $\sqrt{N}$ and variance unity. 
For large $N$ we find that the Gaussian is a good 
fit to the gamma density, for $\vert x-\sqrt{N}\vert << 1$.

In the discussions above on the central limit theorem, we have assumed 
the random variables to be independent and identically distributed.  
The central limit theorem holds well under much more general conditions.

First,  the random variables need not be identical.  To see this, 
let $N_1$ of the $N$ random variables have one distribution with 
say mean $\mu _1 $ and variance $\sigma _1 ^2 $. Let $Y_1$ denote 
their sum. Also let $N_2$ of the $N$ random variables have another 
distribution with mean $\mu_2$ and variance $\sigma _2 ^2$ and let 
their sum be denoted by $Y_2$. We take the limit $N_1 \to\infty$ 
and $N_2\to\infty$ such that $N_1 / N_2 $ remains a constant. 
Asymptotically ($N_1 \to\infty$) the random variable $Y_1$ tends 
to a Gaussian and the characteristic function is 
$\phi_1 (k) = \exp (ikN_1\mu_1 - k^2 N_1 \sigma_1 ^2 /2 )$.  
Similarly the random variable $Y_2$ tends to a Gaussian and the 
characteristic function is 
$\phi_2 (k)= \exp (ikN_2\mu_2 - k^2 N_2 \sigma_2 ^2 /2 )$.  
Their sum $Y=Y_1 + Y_2$ has the characteristic function 
$\phi (k) = \phi _1 (k) \times \phi _2 (k) $, since $Y_1$ and $Y_2$ 
are independent. 
Thus $\phi (k) = \exp \left\{ ik \left( N_1\mu_1 + N_2\mu_2\right) -  
k^2 \left( N_1\sigma_1 ^2 + N_2\sigma _2 ^2 \right) /2\right\}$.  
Fourier inverse of $\phi (k) $ is a Gaussian with mean 
$N_1 \mu _1 + N_2 \mu _2$ and variance 
$N_1 \sigma_1 ^2 +N_2 \sigma_2 ^2$.

Second, the random variables can be weakly dependent, see 
for {\it e.g.} Gardiner \cite{GAR}.  An example is the 
correlated random walk, described in Assignment (6).  
Essentially the adjacent steps of the random walk 
are correlated. Despite this, asymptotically the 
distribution of the position of the random walk is 
Gaussian, see for example \cite{tmj}.

\begin{boxedtext}
\noindent
{\bf Assignment 5}\\

\noindent
Consider a random variable $X$ with probability density $f(x)=\exp (-\sqrt{2}\vert x\vert )/\sqrt{2}$, for $-\infty \le x\le +\infty$.\\
\noindent
(A) What is the characteristic function of $X$~?\\
\noindent
 Let $X_1 , X_2 , \cdots X_N $, be independent and identically distributed random variables with the common density $f(x)$ given above.  Let $Y_N$ be the sum of these. {\it i.e.} $Y_N = \sum_{i=1}^{N} X_i $.\\
\noindent
(B) What is the characteristic function of the random variable $Y_N$~?\\
\noindent 
Let $\bar{Y}_N = Y_N /N $.\\
\noindent
(C) What is the characteristic function of $\bar{Y}_N$~?\\
\noindent
Let $\bar{Y}_{\sqrt{N}}=Y_N / \sqrt{N}$.\\
\noindent
(D-a) What is the characteristic function of $\bar{Y}_{\sqrt{N}}$~?\\
\noindent
(D-b) Taylor expand the characteristic function of $\bar{Y}_{\sqrt{N}}$ in terms of moments and in terms of cumulants.\\
\noindent
(D-c) Employing the Moment and Cumulant expansion, demonstrate the approach to  Gaussian as $N\to\infty$.\\
\noindent
(D-d) What is the mean of the asymptotic Gaussian~?\\
\noindent
(D-e) What is its variance of the asymptotic Gaussian~?
\end{boxedtext}

\begin{boxedtext}
\noindent
{\bf Assignment 6}\\

\noindent
Consider one-dimensional lattice with lattice sites indexed by $(\cdots -3, -2, -1, 0, +1, +2, +3, \cdots)$; a random walk starts at origin and in its first step  jumps to the left site at $-1$ or to the right site $+1$ with equal probability. In all subsequent steps the probability for the random walk to continue in the same direction is $\alpha$ and reverse the direction is $1-\alpha$.  Let $x(n)$ denote the position of the random walk after $n$ steps. Show that asymptotically ($n\to\infty$) the distribution of $x(n)$ is Gaussian. What is the mean of the limiting Gaussian?  What is its variance? See \cite{tmj}
\end{boxedtext}

Thus we find that for the central limit theorem to hold good it is adequate that each of the random variables has a finite variance and thus none of them dominates the sum excessively. They need not be identical; they can also be weakly dependent.

Therefore, whenever a large number of random variables additively determine a parameter, then the parameter tends to have a Gaussian distribution. Upon addition, the individual random variables lose their characters and the sum acquires a Gaussian distribution. It is due to this remarkable fact that Gaussian  enjoys a pre-eminent position in statistics and statistical physics.

Having equipped ourselves with the above preliminaries let us now turn our attention to random and pseudo random numbers that form the backbone of all Monte Carlo simulations.

\noindent 
\section{RANDOM NUMBERS}

\noindent
{\bf What are Random numbers~?}\\

\noindent
We call a sequence of numbers {\it random}, if it is generated by a random physical process.

\noindent
How does randomness arise in a (macroscopic) physical system while its microscopic constituents obey deterministic and time reversal invariant laws~? This issue concerns the search for a microscopic basis for the second law of thermodynamics and the nature and content of entropy(randomness).  I shall not discuss these issues here, and those interested should consult \cite{SECLAW}, and the references therein.

Instead we shall say  that physical processes such as radioactivity decay, thermal noise in electronic devices, cosmic ray arrival times {\it etc.}, give rise to  what we shall call a sequence of truly random numbers. There are however several problems associated with generating random numbers from physical processes.  The major one concerns the elimination of the apparatus bias. See for example \cite{fc} where description of generating truly random numbers from a radioactive alpha particle source is given; also described are the bias removal techniques.

Generate once for all, a fairly long sequence of random numbers from a random physical process. Employ this in all your Monte Carlo calculations. This is a safe procedure.  Indeed tables of random numbers were generated and employed in the early days of Monte Carlo practice, see for example \cite{lhctippet}.  The most comprehensive of them  was published by Rand Corporation \cite{RAND} in mid fifties. The table contained one million random digits.

Tables of random numbers are useful if the  Monte Carlo calculations are carried out manually. However, for computer calculations, use of these tables is impractical.  The reason is simple. A computer has a relatively small internal memory; it cannot hold a large table. One can keep the table of random numbers in an external storage device like a disk or tape; constant retrieval from these peripherals would considerably slow down the computations. Hence, it is often desirable to generate a random number as and when required, employing a simple algorithm that takes very little time and very little memory. This means one should come up with a practical definition of randomness of a sequence of numbers.

At the outset we recognize that there is no proper and satisfactory definition of randomness. Chaitin \cite{chaitin} has made an attempt to capture an intuitive notion of randomness into a precise definition. Following Chaitin, consider the two sequences of binary random digits given below:
\begin{eqnarray}
\begin{array}{cccccccccccccc}
1 & 0 & 1 & 0 & 1 & 0 & 1 & 0 & 1 & 0 & 1 & 0 & 1 & 0\cr
  &   &   &   &   &   &   &   &   &   &   &   &   &  \cr
1 & 1 & 0 & 1 & 0 & 0 & 1 & 0 & 1 & 1 & 0 & 0 & 1 & 1
\end{array}\nonumber
\end{eqnarray}

I am sure you would say that the first is not a random  sequence because 
there is a pattern in it - a repetition of the doublet 1 and 0; the second 
is perhaps a random sequence as there is no recognizable pattern. Chaitin 
goes on to propose that {\bf a sequence of numbers can be called random if 
the smallest algorithm capable of specifying it to the computer has about 
the same number of bits  of information as the sequence itself.}

Let us get back to the two sequences of binary numbers given above.  We recognize that tossing a fair coin fourteen times independently can generate both these sequences. Coin tossing is undoubtedly a truly random process. Also the probability for the first sequence is exactly the same as that for the second and equals  $2^{-14}$, for a fair coin.  How then can we say that the first sequence is not random while the second is~? Is not a segment  with a discernible pattern, part and parcel of an infinitely long  sequence of truly random numbers~?  See \cite{brokendice}
for an exposition of randomness along these lines.

We shall  however take a practical attitude, and consider numbers generated by a deterministic algorithm. These numbers are therefore predictable and reproducible; the algorithm itself occupies very little memory. Hence by no stretch of imagination can they be called random. We shall  call them {\bf pseudo random numbers}.  We shall be content  with pseudo random numbers and employ  them in Monte Carlo calculations. We shall reserve the name pseudo random numbers for those generated by deterministic algorithm and are supposed to be distributed independently and uniformly in the range $0$ to $1$. We shall denote the sequence of pseudo random numbers by $\{ \xi_1 , \xi_2 , \cdots\}$.  For our purpose it is quite adequate if {\it one ensures that the sequence of pseudo random numbers is statistically indistinguishable from a sequence of truly random numbers.} This is a tall order! How can anybody  ensure this~?

\noindent
\subsection{Tests for Randomness}

\noindent
Usually we resort to what are called {\bf tests of randomness}. A test, in a general sense, consists of constructing a function $\psi (r_1, r_2, \cdots )$, where $r_1 , r_2 , \cdots $, are independent variables. Calculate the value of the function for a sequence of  pseudo random numbers by setting $r_i = \xi_i\ \forall \ i=1,\ 2,\ \cdots$.  Compare this value with the value that $\psi$ is expected to have if $\{ r_i : i=1,2,\cdots \}$ were truly random numbers distributed independently and uniformly in the range $0$ to $1$.

For example, the simplest test one can think of, is to set
\begin{equation}
\psi (r_1, r_2, \cdots  )={{1}\over{N}}\sum_{i=1}^{N} r_i ,
\end{equation}
which defines the average of $N$ numbers. For a sequence of $N$ truly random numbers we expect $\psi$ to lie between $.5-\epsilon$ and $.5+\epsilon$ with a certain probability $p(\epsilon)$. Notice that for $N$ large, from the Central Limit Theorem, $\psi$ is Gaussian with mean $0.5$ and variance $\sigma^2 = (12N)^{-1}$. If we take $\epsilon = 2\sigma=1/\sqrt{3N}$, then $p(\epsilon)$ is the area under the Gaussian between $.5-\epsilon$ and $.5+\epsilon$ and is equal to $0.95$. This is called the two-sigma confidence interval. Thus for a sequence of $N$ truly random numbers, we expect its mean to be within $\pm\epsilon$ around $0.5$ with $.95$ probability, for large $N$. If a sequence of $N$ pseudo random numbers has an average that falls outside the interval $(.5-\epsilon , .5+\epsilon)$ then we say that it fails the test at $5\%$ level.

\begin{boxedtext}
\noindent
{\bf Assignment 7}\\

\noindent
Carry out the above test on the random numbers generated by one of the generators in your computer. Does it pass the test at $5\%$ level~?
\end{boxedtext}

Another example consists of setting,
\begin{equation}
\psi (r_1 , r_2 , \cdots )\equiv C(k)  =
{{N \sum_{i=1}^{N} r_i r_{i+k} - \left( \sum_{i=1}^{N} r_i\right)^2}
\over{
N\sum_{i=1}^{N}r_i ^2 - \left( \sum_{i=1}^{N} r_i \right)^2 }},
\end{equation}
with $k=0,1,2,\cdots ,N-1$. For a sequence of truly random numbers, the function  $\psi$ above, which denotes two point auto correlation function, is expected to be unity for $k=0$ and zero for $k\ne 0$.

\begin{boxedtext}
\noindent
{\bf Assignment 8}\\

\noindent
Calculate the auto correlation of the random numbers generated by one of the random number generators in your computer.
\end{boxedtext}

In practice, one employs more complicated tests, by making $\psi$ a more complicated function of its arguments.

An example is the {\bf run test} which is  more sensitive to the correlations.  The idea is to calculate the length of a run of increasing (run-up) or decreasing (run-down) size. We say the run-down   length is $l$ if we have a sequence of random numbers such that $\xi_{l+1} > \xi_l < \xi _{l-1}  < \xi_{l-2} < \cdots \xi_2 < \xi_1$. We can similarly define the  run-up  length.

Let me illustrate the meaning of run-down length by considering a sequence of integers between  $0$ and $9$, given below.
\begin{eqnarray}
\begin{array}{ccccccccccccccccccccccccccc}
 &6  &  &4  &  & 1 &  &3  &  &4 &  &5  &  &3  &  & 2 &  &7  &  &4  & &8 &
\\ &  &  &  &  &  &  &  &  & &  &  &  &  &  &  &  &  &  &  &  &  & \\ | &6
&  &4  &  & 1 &  &3  & | &4 &  &5  & | &3  &  & 2 &  &7  &|  &4  & &8 &|
\\ &  &  &  &  &  &  &  &  & &  &  &  &  &  &  &  &  &  &  &  &  & \\ | &
&  &  &  &  &  &3  & | & &  &1  & | &  &  &  &    &2  &|  &  & & 1 &|
\end{array}\nonumber
\end{eqnarray}
The first row displays the sequence; the second  depicts the same sequence with numbers separated into groups by vertical lines. Each group contains $n+1$ numbers the first $n$ of which are in descending order.  {\it i.e.} these numbers are running down. The descent is broken by the $(n+1)^{{\rm th}}$ number which is greater than the $n^{{\rm th}}$ number. The third row  gives the run-down length, for each group, which is $n$.

\begin{figure}[ht]
\centerline{\psfig{figure=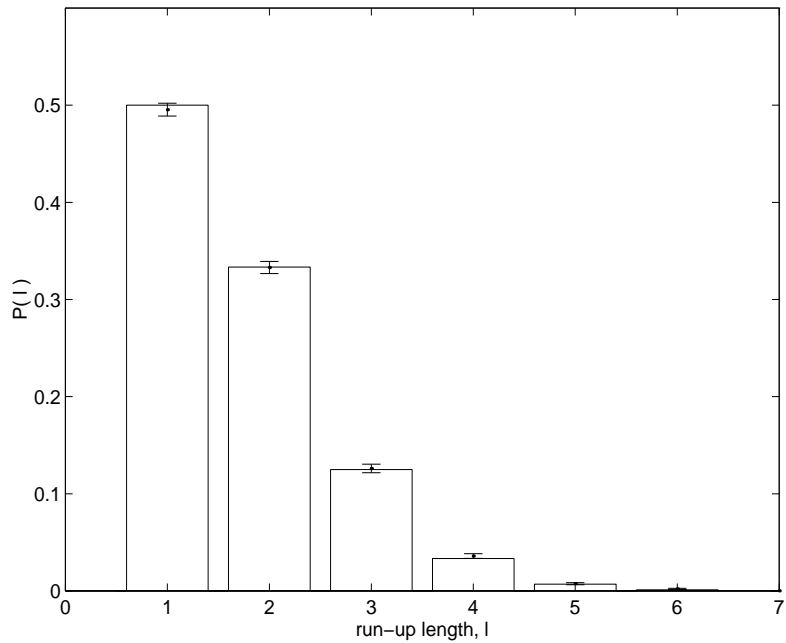,height=8.5cm,width=11.5cm}}
\caption{\protect\small 
Results of a run-up test. The histogram plot corresponds to what is expected from a sequence of truly random numbers. See Eq. (\ref{rundown_eq}). The points with error bars correspond to the ones obtained with random number generator routine RAND of MATLAB  version 4.1. The run-up lengths were generated from a sequence of $10,000$ numbers. The statistical errors in the calculated values correspond to one-sigma confidence interval.}
\label{RUNUP_PS}
\end{figure}

We can calculate the probability distribution of the run-down  length as follows.  Let $P(L\ge l)$ be the probability that the run-down length $L$  is greater than or equal to $l$. To calculate $P(L\ge l)$ we consider a sequence of $l$ distinct random numbers. There are $l!$ ways of arranging these numbers. For a sequence of truly random numbers, each of these arrangements is equally likely. Of these, there is only one arrangement which has all the $l$ numbers in the descending order.  Hence $P(L\ge l) = 1/l!$. Since $P(L= l)= P(L\ge l)- P(L\ge l+1)$, we get
\begin{equation}\label{rundown_eq}
P(l) = {{1}\over{l!}} - {{1}\over{(l+1)!}}\quad.
\end{equation}
Alternately,  for the probability of run-down length  we have,
\begin{eqnarray}
P(L\ge l) & = & \int_{0}^{1}d\xi_1\int_{0}^{\xi_1}d\xi_2 \cdots
\int_{0}^{\xi_{l-1}}d\xi_{l}\nonumber\\
 & = & \int_0 ^1 {{\xi_1 ^{l-1} }\over{(l-1)!}} d\xi_1\nonumber\\ & = &
{{1}\over{l!}}
\end{eqnarray}
In the test, we determine  numerically, the distribution of the run length on the sequence of pseudo numbers and compare it with the exact distribution given by Eq. (\ref{rundown_eq}).  Fig. \ref{RUNUP_PS}
 depicts the results of a run-up  test.  
Description of several of the most common tests for randomness can be found in \cite{TESTS}.

One issue becomes obvious from the above discussions. There is indeed 
an indefinitely large  number of ways of constructing the function 
$\psi$. Hence, in principle a pseudo random number generator can never 
be tested thoroughly for the randomness of the sequence of random numbers 
it generates.  What ever may be the number of tests we employ and 
however complicated they may be, there can be, and always shall  
be, surprises....surprises like the Marsaglia planes discovered 
in the late sixties \cite{MARS} or the parallel lines discovered 
some three years ago by Vattulainen and co-workers \cite{VAT}.  
We shall discuss briefly these issues a bit later.

\begin{boxedtext}
\noindent
{\bf Assignment 9 }\\

\noindent
Carry out  (a) run up and (b) run down tests, on the random numbers generated by one of the random number generators in your computer and compare your results with the exact analytical results. It is adequate if you consider run lengths up to  six or so.
\end{boxedtext}

The struggle to develop better and better random number generators 
and the simultaneous efforts to unravel the hidden order in the 
pseudo random numbers generated, constitute an exciting and continuing 
enterprise, see for example \cite{GUT}. See also Bonelli and Ruffo \cite{BORU} 
for a recent interesting piece of work on this subject.

\noindent
\subsection{Pseudo Random Number Generators}

\noindent
The earliest pseudo random number generator was the {\bf mid-squares} proposed by von Neumann. Start with an $m$ digit integer $\nu_1$. Take the middle $m/2$ digits and call it $\eta_1$. Square $\eta_1$ and call the resulting $m$ digit integer as $\nu_2$. Take the middle $m/2$ digits of $\nu_2$ and call it $\eta_2$. Proceed in the same fashion and obtain a sequence of integers $\eta_1, \eta_2, \cdots$. These are then converted to real numbers between zero and one by dividing each by $10^{m/2}$.  For a properly chosen seed $\nu_1$, this method yields a long sequence of {\it apparently} good random numbers. But on a closer scrutiny, it was found that the mid-squares is not a good random number generator. I shall not spend anymore time on this method since it is not in vogue. Instead we shall turn our attention to the most widely used class of random number generators called  the {\bf linear  congruential generators}, discovered by Lehmer \cite{LEH}. Start with a seed $R_1$ and generate successive integer random numbers by
\begin{equation}\label{cg}
R_{i+1} = a R_i + b\ \ \  (\rm{mod}\  m) ,
\end{equation}
where $a$, $b$, and $m$ are integers. $a$ is called the {\bf generator} or {\bf multiplier}; $b$, the {\bf increment} and $m$ is the {\bf modulus}.

Equation (\ref{cg}) means the following. Start with an integer $ R_1$. This is your choice. Calculate $a\times R_1 + b$. This is an integer. Divide this by $m$ and find the remainder. Call it $R_2$.  Calculate $a\times R_2 + b$ ; divide the result by $m$ ; find the remainder and call it $R_3$. Proceed in the same fashion and calculate the sequence of integers $\{ R_1 , R_2 , R_3, R_4, \cdots\}$, initiated by the seed $R_1$.
Thus $\{ R_i\ :\  i=1,2,\cdots\} $ is a sequence of random integers. The above can be expressed as,
\begin{equation}
R_{i+1} = (a\times R_i + b ) - \left[ {{ (a\times R_i ) +b
}\over{m}}\right]\times m ,
\end{equation}
where the symbol $[\  \eta\  ]$ represents the largest integer less than or equal to $\eta$; {\it e.g.}, $[ \pi ]=3,\  [\sqrt{2}]=1,\ [ 4.999]=4,\ ${\it etc.}.  The random integers $\{ R_j \}$ can be converted to floating point numbers by dividing each by $m$. {\it i.e.}, $\xi _j = R_j /m$. Then $\{ \xi_i : i=1,2,\cdots\}$ is the desired sequence of  pseudo random numbers in the range $0$ to $1$.

The linear congruential generator is robust (over forty years of heavy-duty use!), simple and fast; the theory is reasonably well understood. It is undoubtedly an excellent technique and gives fairly long sequence of reasonably good random numbers, when properly used. We must exercise great care in the  use of congruential generators, lest we should fall into the {\it deterministic } traps.  Let me  illustrate:

Consider Eq. (\ref{cg}). Let us take $a=5,\ b=1,\ R_1 =1,$ and $m=100$. The results of the recursions are shown in Table (3).  

\begin{table}[h]\label{lcga5b1m100_tab}
\caption{\protect\small  Linear congruential recursion with $a=5$, $b=1$, $m=100$, and 
$R_1 = 1$.}
\bigskip
\hrule
\bigskip
\begin{tabular}{lccrllrrr}
     &   &          &    &     &                 &      &              &
\\ $R_1$&   &          &    &     &                 &      & $   =$
&  $1$ \\ &   &          &    &     &                 &      &
&     \\ $R_2$&$=$&$(5\times$&$1 $&$+1)$&(mod $100)=$&$6  $ &(mod
$100)=$&$6 $ \\ &   &          &    &     &                &    &
&     \\ $R_3$&$=$&$(5\times$&$6$ &$+1)$&(mod $100)=$&$31 $&(mod
$100)=$&$31$ \\ &   &          &    &     &                &     &
&     \\ $R_4$&$=$&$(5\times$&$31$&$+1)$&(mod $100)=$&$156$&(mod $100)=$&
$56$\\ &   &          &    &     &                &     &                &
\\ $R_5$&$=$&$(5\times$&$56$&$+1)$&(mod $100)=$&$281$&(mod $100)=$& $81$\\
&   &          &    &     &                &     &                &     \\
&   &          &    &     &                &     &                &     \\
$R_6$&$=$&$(5\times$&$81$&$+1)$&(mod $100)=$&$406$&(mod $100 )=$&$6  $\\ &
&          &    &     &                &     &                &     \\
\end{tabular}
\bigskip
\hrule 
\end{table}

We see from Table (3) 
 that $R_6 = R_2 ; R_7
= R_3 , \cdots $, and the cycle repeats.  The period is four. We just get
four random(?) integers!

Consider another example with $a=5,\ b=0,\ m=100, $ and $R_1 = 1$.  Table (4) 
 depicts the results of the linear congruential recursion.

\begin{table}[h]\label{lcga5b0m100_tab} 
\caption{\protect\small Linear congruential recursion with $a=5$, $b=0$, $m=100$, and $R_1 = 1$}
\bigskip
\hrule
\begin{tabular}{lccrllrllr}
     &   &          &    &               &   &    &                &    &
\\ $R_1$&   &          &    &               &   &    &                &$=$
& $1$\\ &   &          &    &               &   &    &                &
&    \\ $R_2$&$=$&$(5\times$&$1)$&(mod $100)$ &$=$&$5 $&(mod $100)$ &$=$ &
$5$\\ &   &          &    &               &   &    &               &   &
\\ $R_3$&$=$&$(5\times$&$5)$&(mod $100)$ &$=$&$25$&(mod $100)$&$=$&$25$\\
&   &          &    &               &   &    &                &   &    \\
$R_4$&$=$&$(5\times$&$25)$&(mod $100)$ &$=$&$125$&(mod $100)$&$=$&$25$\\ &
&          &    &               &   &    &                &   &  \\
\end{tabular}
\bigskip
\hrule
\end{table}

Starting with the seed $1$, we get $5$ followed by  an endless array of
$25$, as seen from Table (4). 
These
examples illustrate dramatically how important it is that we make a proper
choice of $a$, $b$ and $m$ for decent results.
   
Clearly, whatever be the choice of $a$, $b$ and $m$, the sequence of
pseudo random integers generated by the linear congruential technique
shall repeat itself after utmost $m$ steps. The sequence is therefore
periodic. Hence, in applications we must ensure that the number of random
numbers required for any single simulation must be much less than the
period.  Usually $m$ is taken very large to permit this.

For the linear  congruential generator, we can always get a sequence with
full period, $m$, if we ensure:
\\

1.  $m$ and $b$ are relatively prime to each other; {\it i.e.}\
$\gcd(m,b)=1: $
\\

2.  $a \equiv\  1\  (\rm{mod}\  p)$ for every prime factor $p$ of $m$; and
\\

3.  $a\equiv\  1\  (\rm{mod}\ 4)$, if $m\equiv\  0\  \rm{(mod}\  4)$. \\

For example, let $a=7$, $b=13$ and $m=18$. Check that this choice
satisfies the above conditions. The results of the linear congruential
recursion with the above parameters are depicted in Table (5).  

\begin{table}[h]\label{lcga7b13m18_tab}
\caption{\protect\small  Linear congruential recursion with $a=7$, $b=13$,$m=18$ and $R_1=1$}
\bigskip
\hrule 
\bigskip
\begin{eqnarray}
\begin{array}{lclrllcrr}
R_1    &   &          &   &                &   &     &            = & 1 \\
R_2    &=& (7\times &  1&   +13) & (\rm{mod}\  18)  =&  20  & (\rm{mod}\
18) =& 2\\ R_3    &=& (7\times & 2 &  +13) & (\rm{mod}\  18)  =&  27  &
(\rm{mod}\  18) =& 9\\ R_4    &=& (7\times & 9 &  +13) & (\rm{mod}\  18)
=&  76  & (\rm{mod}\  18) =& 4\\ R_5    &=& (7\times & 4 &  +13) &
(\rm{mod}\  18)  =&  41  & (\rm{mod}\  18) =& 5\\ R_6    &=& (7\times & 5
&  +13) & (\rm{mod}\  18)  =&  48  & (\rm{mod}\  18) =& 12\\ R_7    &=&
(7\times & 12 &   +13) & (\rm{mod}\  18)  =&  97  & (\rm{mod}\  18) =& 7\\
R_8    &=& (7\times & 7  & +13) & (\rm{mod}\  18)  =&  62  & (\rm{mod}\
18) =& 8\\ R_9    &=& (7\times & 8  & +13) & (\rm{mod}\  18)  =&  69  &
(\rm{mod}\  18) =& 15\\ R_{10} &=& (7\times & 15 &  +13) & (\rm{mod}\  18)
=& 118  & (\rm{mod}\  18) =& 10\\ R_{11} &=& (7\times & 10 &  +13) &
(\rm{mod}\  18)  =&  83  & (\rm{mod}\  18) =& 11\\ R_{12} &=& (7\times &
11 &  +13) & (\rm{mod}\  18)  =&  90  & (\rm{mod}\  18) =& 0\\ R_{13} &=&
(7\times & 0  & +13) & (\rm{mod}\  18)  =&  13  & (\rm{mod}\  18) =& 13\\
R_{14} &=& (7\times & 13 &  +13) & (\rm{mod}\  18)  =& 104  & (\rm{mod}\
18) =& 14\\ R_{15} &=& (7\times & 14 &  +13) & (\rm{mod}\  18)  =& 111  &
(\rm{mod}\  18) =& 3\\ R_{16} &=& (7\times & 3  & +13) & (\rm{mod}\  18)
=&  34  & (\rm{mod}\  18) =& 16\\ R_{17} &=& (7\times & 16 &  +13) &
(\rm{mod}\  18)  =& 125  & (\rm{mod}\  18) =& 17\\ R_{18} &=& (7\times &
17 &  +13) & (\rm{mod}\  18)  =& 132  & (\rm{mod}\  18) =& 6\\ & &
&            &                   &      &                  &   \\ R_{19}
&=& (7\times & 6  & +13) & (\rm{mod}\  18)  =&  55  & (\rm{mod}\  18) =& 1
\end{array}\nonumber
\end{eqnarray}
\bigskip
\hrule 
\end{table}

Thus we get the sequence $$
\{ 1,2,9,4,5,12,7,8,15,10,11,0,13,14,3,16,17,6\}
$$ of eighteen distinct integers between $0$ and $17$, with full period
$18$. Divide each number in this sequence by $18$ and get a sequence of
real numbers between $0$ and $1$.  That the period is maximum $m$ only
ensures that the numbers are uniform in the range $0$ to $1$.  The numbers
may be correlated or may have a hidden order.

\begin{figure}[ht]
\centerline{\psfig{figure=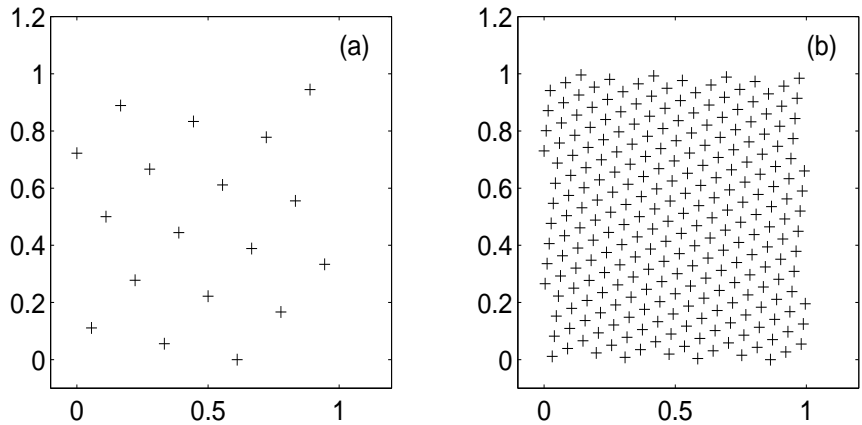,height=5.8cm,width=11.5cm}}
\caption{Random numbers of the linear congruential generator embedded
in a two dimensional phase space. (a) depicts the results for the
generator with $a=7$, $b=13$  and $m=18$. (b) depicts the results for the
generator with $a=137, b=187$ and $m=256$. (Both these generators  give
full period)}
\label{MARSAGLIA_PS}
\end{figure}

Let us embed the above sequence in a two dimensional phase space. This is
carried out as follows. Form two-dimensional embedding vectors: $(\xi_1
,\xi_2)$, $(\xi_2 , \xi_3 )$, $\cdots$, $(\xi_m , \xi_1 )$. We have thus
$m$ vectors and in our example $m=18$. Each of these vectors can be
represented by a point in two dimensional phase space.  
Fig. \ref{MARSAGLIA_PS}a depicts these eighteen  points.  We observe that the
points fall neatly on parallel lines.

Take another example with $a=137$, $b=187$, and $m=256$. We get a sequence
of $256$ distinct integers between $0$ and  $255$ in some order starting
from the chosen seed. Let us convert  the numbers into real numbers between zero and one by dividing each by 256.  Let $\{ \xi_i :i=1,2,\cdots , 256\}$ denote the sequence. Embed the sequence in a two-dimensional \ \lq phase space\rq\  , by forming two-dimensional vectors as discussed above. 
Fig. \ref{MARSAGLIA_PS}b depicts the results of this exercise. 
The vectors clearly fall on parallel lines.

That the random numbers generated by linear congruential generators form neat patterns when embedded in two and higher dimensional phase space was known since as early as the beginning of sixties, see for example \cite{greenberger}. But no one recognized this as an inherent flaw in the linear congruential generators. It was generally thought that if one takes a linear congruential generator with a very long period, one would not perhaps  see any  patterns \cite{chambers}. 

Long periods can be obtained by choosing the modulus $m$  large  and by choosing appropriate values of the multiplier $a$ and the increment $b$ to get the full period.

\begin{boxedtext}
\noindent
{\bf Assignment 10}\\

\noindent
Construct a linear congruential  generator that gives rise to a long sequence of pseudo random numbers with full period.  Embed them in two or higher dimensional phase space and see if there are any patterns.  Test for the randomness of the sequence.
\end{boxedtext}

The modulus $m$ is usually taken as $2^{t-1}-1$, where $t$ is the number of bits used to store an integer and hence is machine specific. One of the $t$ bits is used up for storing the sign of the integer. The choice $a=7^{5}=16801$, $b=0$ and $m=2^{31}-1$, for a $32$ bit machine has been shown to yield good results, see \cite{PTVF}. The Ahren generator which specifies $a=2^{t-2}(\sqrt{5}-1)/2$, $b=0$, has also been shown to yield \ \lq good\rq\   random numbers.  The linear  congruential generators have been successful and invariably most of the present day pseudo random number generators are of this class.

In the late sixties, Marsaglia \cite{MARS} established unambiguously that the formation of lattice structure is an inherent feature of a linear congruential generator:
\\

\noindent
{\bf If $d$-tuples $(\xi_1 ,\ \xi_2,\ \cdots ,\ \xi_d )$, $(\xi_2 ,\ \xi_3,\ \cdots ,\ \xi_{d+1} )$, $(\xi_3 ,\ \xi_4,\ \cdots ,\ \xi_{d+2} )$, $\cdots$, of the random numbers produced by linear congruential generators are viewed as points in the unit hyper cube of $d$ dimensions, then all the points will be found to lie in a relatively small number of parallel hyper planes. Such structures occur with any near congruential generator and in any dimension.}
\\

The number of such {\bf Marsaglia planes} does not exceed $(d!2^t)^{1/d}$. Besides the points on several Marsaglia planes form regular patterns.  Thus it is clear now that the existence of Marsaglia planes  is undoubtedly {\bf a serious defect} inherent in the linear congruential generators.

Then there is the other matter of the presence of correlations, hopefully weak, in the sequence of pseudo random numbers. If your Monte Carlo algorithm is sensitive to the subtle correlations present in the sequence, then you are in problem. For example Ferrenberg, Landau and Wong \cite{flj}, found that the so-called high quality random number generators led to subtle but dramatic errors in algorithm that are sensitive to the correlations. See also \cite{badrn}.

One should perhaps conclude  that the {\bf linear congruential generators are not suitable for Monte Carlo work!} Before we come to such a harsh conclusion, let us look at the issues in perspective. If your Monte Carlo program does not require more than a small  fraction (smaller the better) of the full period of the random numbers, then most likely, the presence of Marsaglia lattice structure will  not affect your results. All the linear congruential generators in use today have long periods. Also one can think of devices to increase the number of Marsaglia planes. For example the Dieter-Ahren \ \lq solution\rq\   to the Marsaglia problem is the following algorithm,
\begin{equation}
R_{i+2} = a R_{i+1}  + b R_i \quad\quad (\rm{mod}\  m),
\end{equation}
which requires two seeds. The modulus $m=2^{t-1}-1$. For a proper choice of $a$ and $b$, the number of Marsaglia planes can be increased by a factor of $2^{t/d}$.

Thus over the period of thirty  years we have learnt to live with the lattice defect of the linear congruential generators. However at any time you get the feeling that there is something wrong with the simulation results and you   suspect that this is caused by  the Marsaglia lattice structure of the linear congruential generator then you should think of employing some other random number generator,  like the {\bf inversive congruential generator (ICG)} proposed in 1986 by Eichenauer and Lehn \cite{eichenauer} or the {\bf explicit inversive congruential generator (EICG)}, proposed in 1993 by Eichenauer-Hermann \cite{eichher}. Both these generators do not suffer from the lattice or hyper plane structure defect. I shall not get into the details of these new inversive generators, except to say that the inversive congruential generator  also employs recursion just as  linear congruential generators do, but the recursion is based on a nonlinear function.

More recently in the year 1995, Vattulainen and co-workers \cite{VAT} found, in the context of linear congruential generators,  that the successive pairs $(\xi_t , \xi_{t+l})$ exhibit a pattern of parallel lines on a unit square. This pattern is observed for all $(\xi_t , \xi_{t+l})$ with $l=1,2,\cdots $, where the modulus $m=2^{31}-1$ for a 32 bit machine. The number of parallel lines strongly depends on $l$. For $l=(m-1)/2$ the pairs fall on a single line, and for $l=\{ (m-1)/2\} \pm 1$, the pairs fall on so large a number of parallel lines that they can be considered as space filling. This curious phenomenon of switching from pairs falling on one or a few parallel lines to pairs falling on several lines upon tuning the parameter $l$ is termed as transition from regular behaviour to chaotic behaviour, see \cite{MATTIS}. Thus there seems to be a connection between pseudo random number generators and chaotic maps.  {\bf Chaos}, we know, permits only short time predictability; no long term forecasting is possible.  The predictability gets lost at long times.....at times greater than the inverse of the largest Lyapunov exponent. It is rather difficult to distinguish a chaotic behaviour from a random behaviour.  Thus Chaos can provide an effective source of randomness. This is definitely meaningful idea since chaotic and random behaviours have many things common, see for example \cite{luscher}, where Chaos has been proposed as a source of pseudo random numbers.  But then we know there is an order, strange (fractal) or otherwise, in Chaos.  What exactly is the connection between the Marsaglia order found in the context of linear congruential generators and the fractal order in Chaos~? If answers to these and
several related questions can be found, then perhaps we can obtain some insight into the otherwise occult art of pseudo random number generators. A safe practice is the following. Consider the situation when you are planning to use a standard Monte Carlo code on a new problem; or consider the situation when you are developing a new Monte Carlo algorithm. In either case, carefully test your Monte Carlo along with the random number generator, on several standard problems for which the results are
reasonably well known.  This you should do irrespective of how \ \lq famous\rq\   the random generator is, and how many randomness tests it has been put through already.   Afterall your Monte Carlo programme itself can be thought of as a new test of randomness of the pseudo random number generator.

I shall stop here the discussion on pseudo random number generators 
and get on to Monte Carlo. In the next section I shall discuss 
random sampling techniques that transform the pseudo random 
numbers, $\{ \xi _i \}$, independent and uniformly distributed 
in the range $0$ to $1$,  to $\{ x_i\}$, 
having the desired distribution in the desired range.

\noindent
\section{RANDOM SAMPLING TECHNIQUES}

\noindent
The {\bf random sampling techniques} help us convert a sequence of random numbers $\{ \xi_i \}$, uniformly distributed in the interval $(0,1)$ to a sequence $\{ x_i \}$ having the desired density, say $f(x)$.  There are several techniques that do this conversion. These are direct inversion, rejection methods, transformations, composition, sums, products, ratios, table-look-up of the cumulative density with interpolation, construction of equi-probability table, and a whole host of other
techniques. In what follows we shall illustrate random sampling by considering in detail a few  of  these techniques.

\noindent
\subsection{Inversion Technique}

\begin{figure}[ht]
\centerline{\psfig{figure=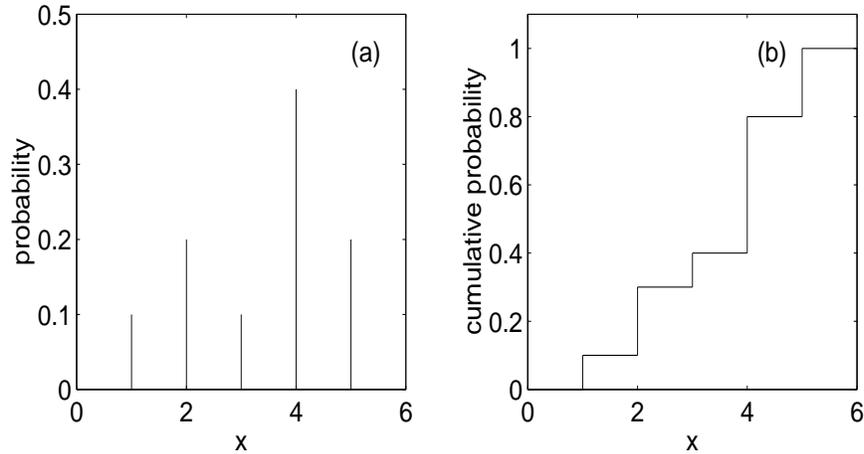,height=6.2cm,width=11.5cm}}
\caption{\protect\small Discrete distribution. (a) probabilities (b) cumulative 
probabilities}
\label{DISCINV_PS}
\end{figure}

\noindent
The simplest is based on the direct {\bf inversion} of the cumulative density function of the random variable $X$.  We shall first consider sampling from a discrete distribution employing inversion technique. Let $\{ p_i\ :\ i=1,2,\cdots ,N\}$ be the discrete probabilities for the random variable $X$ to take values $x_i(=i)$. Fig. \ref{DISCINV_PS}a depicts an example with $N=5$. We first construct the cumulative probabilities $\{P_i\ :\  i=0,N\}$ where $P_0 = 0;\  P_1 = p_1;\ P_2 = p_1 + p_2 : \cdots P_k =p_1 + p_2 + \cdots p_k;\  P_N =1$. 
Fig. \ref{DISCINV_PS}b depicts the cumulative probabilities as staircase  of non-decreasing heights. The procedure is simple. Generate $\xi$, a random number uniformly distributed in the range $(0,1)$.  Find $k$ for which $P_{k-1} \le \xi < P_k$. Then $x_k (=k)$ is the sampled value of $X$. This is equivalent to the following.  The value of $\xi$ defines a  point on the
$y$ axis of Fig. \ref{DISCINV_PS}b between zero and one. Draw a line parallel to the $x$ axis at $(0,\xi)$. Find the vertical line segment intersected by this line.  The vertical line segment is then extended downwards to cut the $x$ axis at say $x_1$. Then $x_1$ is the (desired) sampled value of $X$.

\begin{boxedtext}
\noindent
{\bf Assignment  11}\\

\noindent
Employing inversion technique sample from (a) binomial and (b) Poisson distributions.  Compare the frequency distribution you obtain with the exact.
\end{boxedtext}
The above can be generalized to continuous distribution. We have the cumulative probability distribution $F(x)=\int_{-\infty}^{x}f(x')\ dx'$. Note that $F(-\infty)=0$ and $F(+\infty)=1$. Given a random number $\xi_i$, we have $x_i = F^{-1}(\xi_i)$.

\begin{figure}[ht]
\centerline{\psfig{figure=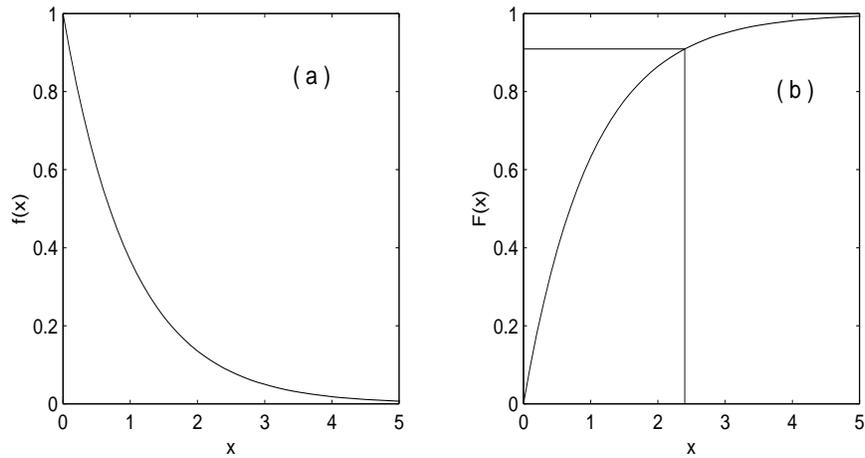,height=6.2cm,width=11.5cm}}
\caption{\protect\small (a) exponential distribution  (b) cumulative
distribution}
\label{EXPINV_PS}
\end{figure}

An example with exponential distribution is shown in Fig. \ref{EXPINV_PS}. We have plotted in Fig. \ref{EXPINV_PS}a the exponential distribution for $x\ge 0$. 
Fig. \ref{EXPINV_PS}b depicts the cumulative distribution. Select a point on the $y$ axis of Fig. \ref{EXPINV_PS}b randomly between zero and one.  Draw a line parallel to $x$ axis passing through this point. Find where it intersects the curve $F(x)$. Read off the $x$ coordinate of the intersection point, and call this $x_1$. Repeat the above several times and get $\{ x_i\  :\ i=1,2,\cdots\}$. These numbers will be exponentially distributed.

\begin{boxedtext}
\noindent 
{\bf Assignment 12}\\

\noindent 
Devise a technique to sample from the distribution $f(x)=\exp (-\sqrt{2}\vert x\vert )/\sqrt{2}\ {\rm for } -\infty \le x \le +\infty$. Generate $\{ x_i : i=1,2,\cdots ,N\} $.  Sum up these numbers and divide by $\sqrt{N}$. Call this $y$. Generate several values of $y$ and plot their frequency distribution in the form of a histogram. Carry out this exercise with $N=1,2,\cdots$ and demonstrate the approach to Gaussian as $N$ becomes larger and larger.
\end{boxedtext}

In fact for the exponential distribution we can carry out the inversion analytically. We set $\xi = F(x)$. It is clear that the number $F(x)$ is uniformly distributed between $0$ and $1$. Hence the probability that it falls between $F(x)$ and $F(x)+dF(x)$ is $dF(x)$, which is equal to $f(x)dx$. Hence $x=F^{-1}(\xi)$ is distributed as per $f(x)$. For exponential distribution we have $F(x)=1-\exp (-x)$. Hence $x=-\log (1-\xi)$. Since $1-\xi$ is also uniformly distributed we can set $x=-\log (\xi)$.

\begin{boxedtext}
\noindent
{\bf Assignment 13}\\

\noindent
Generate $N$ independent random numbers from an exponential distribution. Sum them  up  and divide  by $\sqrt{N}$; call the result  $y$.  Generate a large number of values of $y$ and plot their frequency distribution.  Plot on the same graph the corresponding gamma density and Gaussian and compare.
\end{boxedtext}

\begin{boxedtext}
\noindent
{\bf Assignment 14}\\

\noindent
Start with $N$ particles, indexed by  integers $i=1,2,\cdots , N$.  ({\it e.g.} $N=1024$). Initialize  $x(j)=\lambda\ \forall\ j=1,2,\cdots ,N$, where $\lambda $ is the desired exponential decay constant. ({\it e.g.,} $\lambda =1$). The algorithm conserves $\sum_{i=1}^{N}x(i) = N\lambda$. Select independently and randomly two particles, say with indices $i$ and $j$, and $i\ne j$. Let $S= x(i) + x(j)$.  Split $S$ randomly into two parts. Set $x(i)$ to one part and $x(j)$ to the other part.  Repeat the above for a warm up time of say $4\times N$ iterations.  Then every subsequent time you select two particles ($k$ and $l$), the corresponding $x(k)$ and $x(l)$ are two independent random numbers with exponential distribution:\   $\lambda\exp (-\lambda x)\ {\rm for}\ 0\le x\le \infty$.
\\
(a) Implement the above algorithm and generate a large number of random numbers. Plot their frequency distribution and check if they follow exponential distribution.\\ (b) Prove analytically that the above algorithm leads to independent and exponentially distributed random numbers in the limit $N\to\infty$.
\end{boxedtext}.

Analytic inversion to generate exponentially distributed random numbers is not necessarily the most 
robust and fast of the techniques.  There are several alternate procedures for sampling from 
exponential distribution without involving logarithmic transformation, see for example \cite{sampexp}.  
The subject of developing ingenious algorithms for generating exponentially distributed random numbers 
continues to attract the attention of the Monte Carlo theorists.  For example, recently 
Ferna\'ndez and Rivero \cite{ferriv}  have proposed a simple algorithm to generate independent and 
exponentially distributed random numbers.  They consider a collection of $N$ particles each having 
a certain quantity of energy to start with.  Two distinct particles are chosen at random; their energies 
are added up; the sum is divided randomly into  two portions and assigned to the two particles. 
When this procedure is repeated several times, called the warming up time, the distribution of 
energies amongst the particles becomes exponential. Note that this algorithm conserves the total 
energy.  It has been shown  that about one thousand particles
are adequate; the warming up time is $4N$ or so. For details 
see \cite{ferriv}; see also \cite{wallace}.

\noindent
\subsection{Rejection  Technique}

\noindent
Another useful random sampling technique is the so called {\bf rejection
technique}. The basic idea is simple. From a set of random numbers discard
those that do not follow the desired distribution.  What is left out must
be distributed the way we want.

Let $f(x)$ be the desired distribution, defined over the range $(\alpha , \beta)$. First select a suitable bounding function $g(x)$ such that $C\times g(x) \ge f(x)$ for all values of $\alpha \le x \le \beta$. Also $g(x)$ much be such that it is easy to sample from.  Sample a set of $\{ x_i : i=1,2,\cdots , N\}$ independently and randomly from $g(x)$. For each value of $x_i$ select randomly a number between $0$ and $C\times g(x_i)$. Call this set $\{ y(x_i)\}$.  From the set $\{ x_i : i=1,2,\cdots ,N\}$ discard those for which $y(x_i) > f(x_i)$.  The remaining $x_i$-s shall have the desired distribution $f(x)$.  The efficiency of a rejection technique is the percentage of the attempts that get accepted and is given by the inverse of the area under the curve $C\times g(x)$. Usually $g(x)$ is taken as a constant for all $x$ and $C$ is the maximum value that $f(x)$ takes. The rejection technique would be inefficient if $C\times  g(x)$ and $f(x)$ do not have similar shapes and ranges.

\begin{figure}[ht]
\centerline{\psfig{figure=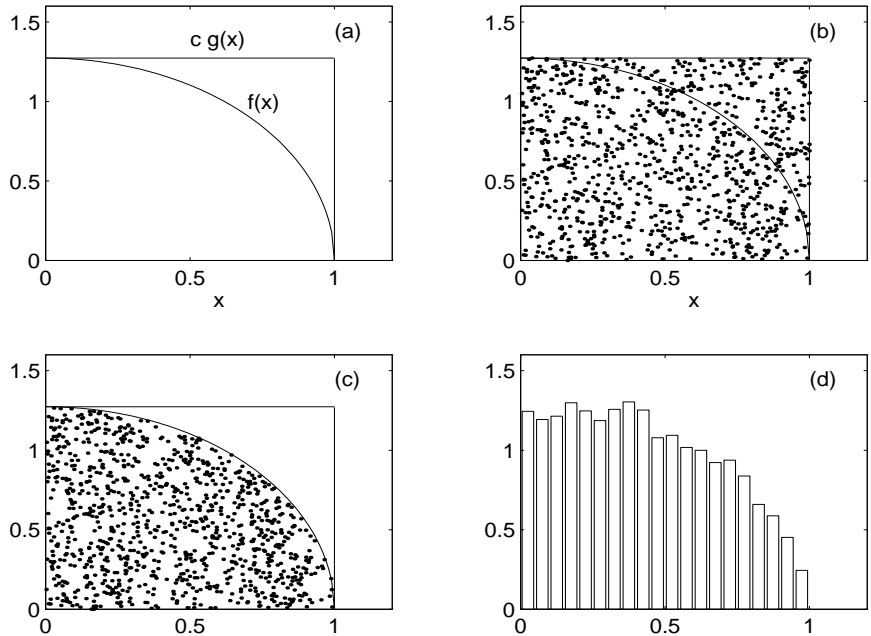,height=8.5cm,width=11.5cm
}}
\caption{\protect\small Illustration of the rejection technique. 
(a) the circular density and the $C\times $ the bounding function. (b)
random points that uniformly fill up the rectangular box. (c) points that
get accepted in the sampling procedure. (d) the distribution of the
accepted values of $x$.}
\label{REJECTION_PS}
\end{figure}

Let me illustrate the rejection technique by considering the circular
probability density function
\begin{equation}
f(x)={{4}\over{\pi}}\sqrt{1-x^2}\quad {\rm for} \quad 0\le x\le 1.
\end{equation}
A convenient  choice of the bounding function is $g(x)=1\ \forall \
\quad 0\le x\le 1$. Thus sampling from $g(x)$ is equivalent to setting
$x_i= \xi _i$. The value of $C$ is $4/ \pi$. Fig. \ref{REJECTION_PS}a
depicts the density function $f(x)$ and the function $C\times g(x)$.
Generate a pair of random numbers $( \xi_i , \xi _j )$ from $U(0,1)$.  Set
$x_i = \xi _i$. Calculate $f_i = f(x_i)$ and $y_i = C\times \xi_j$.  If
$y_i \le f_i$, accept $x_i$.  Repeat the above procedure several times.
Fig. \ref{REJECTION_PS}b depicts $\{ (x_i,y_i);i=1,2,\cdots ,1000\}$ and these
points are seen to fill up the rectangle bounded by the lines $y=0,
y=4/\pi, x=0$ and $x=1$.  Fig. \ref{REJECTION_PS}c depicts the
accepted pairs $\{ (x_i, y_i)\}$.  Fig. \ref{REJECTION_PS}d depicts
the histogram of the distribution of the accepted values of $x_i$.

In the introduction, I mentioned of  the \ \lq pebble-throwing\rq\ game
popular in the province of Monacco, from which the name Monte Carlo came.
What has been described above is a straight forward adaptation of the
same (Monte Carlo) game to sampling from a (normalized) distribution. I
said earlier that one can make an estimate of the value of $\pi$ from the
Monte Carlo game. How do we do this~?

\begin{figure}[ht]
\centerline{\psfig{figure=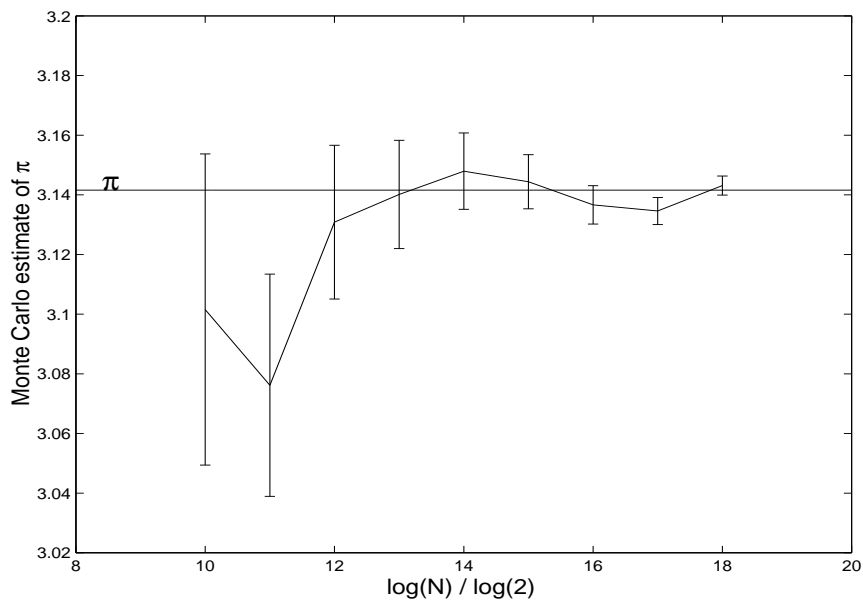,height=8.5cm,width=11.5cm}}
\caption{\protect\small Monte Carlo estimate of $\pi$ employing
rejection technique}
\label{pi_ps}
\end{figure}

In the rejection technique described above,  we sample a point randomly
from the rectangular region, see Fig. \ref{REJECTION_PS}a. In the next
step we either accept the point or reject it. Thus there are only two
outcomes to the experiment.  The point either falls inside the \ \lq
circular distribution\rq\   curve (and hence gets accepted) with a
probability $p=\pi /4$ or outside (and hence gets rejected)  with a
probability $q=1-p=1-\pi /4$.  We are essentially  tossing  a loaded coin;
the probability of Heads is $p$ and of Tails $q=1-p$. Let $n$ denote the
number of Heads in $N$ independent throws of a loaded coin.  The
distribution of the random variable $n$ is Binomial.  The quantity of
interest to us is $n/N$, whose mean and standard deviation can easily be
calculated, and are given by $p$ and $\sqrt{p(1-p)}/\sqrt{N}$
respectively. Thus we say the Monte Carlo estimate of $\pi$ is $4n/N \pm
(4/N^{3/2})\sqrt{n(N-n)}$, where the error term is the one-sigma
confidence interval. Fig.  \ref{pi_ps} depicts the estimated value of
$\pi$ with the one-sigma confidence error bar,   as a function of the
logarithm (to the base 2) of the  number of trials $N=2^{10} , 2^{11},
\cdots 2^{18}$. The statistical convergence to the right answer as the
sample size increases is seen.

\begin{boxedtext}
\noindent
{\bf Assignment 15}\\

\noindent
Employ rejection technique to sample from
$f(x)={{2}\over{\pi}}\sqrt{1-x^2}\ \ \ \rm{for} -1 \le  x \le  1$.  Plot
the distribution of the accepted sequence of numbers in the form of a
histogram and compare with the exact.  Calculate the value of $\pi$ from
the above experiment~? What is the statistical error~?
\end{boxedtext}

\noindent
\subsection{Sampling from a  Gaussian} 

\noindent
The Gaussian is the most important distribution in statistics and statistical physics. It is also one of the {\it  richest}, in the sense, a very large number of techniques have been proposed to generate Gaussian random numbers. The simplest perhaps is the one that makes use of the Central Limit Theorem: take the sum of $N$ random numbers, uniformly distributed in the range $(0,1)$. Let $Y _N$ denote the sum. By Central Limit Theorem, the distribution of $Y_N$ will tend to a Gaussian in the limit $N\to\infty$. The mean of the Gaussian is $N/2$ and its variance is $N/12$.  Usually one wants a standard Gaussian with mean zero and variance unity. Therefore we define the quantity $q$, 
\begin{equation}
q={{Y _N -(N/2)}\over{\sqrt{N/12}}},
\end{equation}
which has a Gaussian  distribution of zero mean and unit variance for large $N$. A convenient choice is $N=12$, which reduces  $q$ to $Y_{12} -6$: Take twelve pseudo random numbers and add them up;  subtract six from the sum and you have a  mean zero variance unity Gaussian random number. This method of generating Gaussian random numbers is not exact since the tails are absent. Note that the value of $q$ is restricted to the interval $(-6,+6)$.

There is a clever technique that transforms two independent uniform random variables $\xi_1$ and $\xi_2$ to two independent Gaussian random variables $q_1$ and $q_2$, as follows.
\begin{eqnarray}
q_1 & = & \sqrt{-2\  \ln \  \xi_1 }\ \cos (2\pi \xi_2) ,\nonumber\\ q_2 &
= & \sqrt{-2\  \ln \  \xi_1 }\ \sin (2\pi \xi_2) .
\end{eqnarray}
This method, called {\bf Box-Muller algorithm} \cite{bm}, is easy to program. This algorithm is quite popular. However it has a few drawbacks. Box-Muller algorithm  uses several multiplication, one logarithmic function, one trigonometric function and one square root function; hence it is rather slow.  Besides the tails differ markedly from the true when a linear congruential generator is used.

  Muller \cite{muller} gives an account of several methods of sampling from a Gaussian distribution.

\begin{boxedtext}
\noindent
{\bf Assignment 16}\\

\noindent
Generate $N$ random numbers from $U(0,1)$. Add them up and subtract $N/2$. Divide the result by $\sqrt{N/12}$. Call this $y$. Generate a large number of $y$ and plot their frequency distribution. Take $N=2,3,4,\cdots$, and demonstrate the approach to Gaussian (of mean zero and variance unity).
\end{boxedtext}

Recently Ferna\'ndez and Criado \cite{fercri} have proposed 
a fast algorithm to generate Gaussian random numbers.  
Their algorithm is based on an $N$ particles closed system 
interacting two at a time, conserving energy. Start with 
$N$ particles each having the same velocity unity. 
{\it i.e.} $\{ v_i =1\ \forall\ i=1,2,\cdots ,N\} $.  Pick up two 
particles at random; let them be $i$ and $j$, and $i\ne j$.  
Reset their velocities as per the following iteration rule,
\begin{eqnarray} 
v_i ({\rm new} ) & = & {{ v_j ( {\rm old} ) + v_i ( {\rm old})}\over{
\sqrt{2} }}\nonumber\\ v_j ( {\rm new} )& = & {{ v_j ( {\rm old} ) - v_i (
{\rm old})}\over{ \sqrt{2} }}.
\end{eqnarray}
   
Repeat the above several times.  After initial {\it warm up time} 
of say $4N$ iterations or so, the velocities of the pair of  
particles you  pick up in all subsequent iterations  are the 
desired pairs of independent Gaussian random numbers with 
mean zero and variance unity.   Gaussian of desired mean, 
say $\mu$ and standard deviation, say $\sigma$ can be 
obtained by the transformation $x=\sigma v + \mu$.  Note 
that $\sum_{i=1}^{N}v_i ^2 = N$ at all stages of iteration.  
This algorithm is found  to be ten times faster than 
the Box-Muller algorithm. For most applications, it is 
adequate if ten thousand to hundred thousand particles 
are considered, see \cite{fercri} for details.

\begin{boxedtext}
\noindent
{\bf Assignment 17}\\

\noindent
(a) Prove analytically  that the Ferna\'ndez-Criato algorithm leads to independent Gaussian random numbers. Prove that the sampling is ergodic. See \cite{fercri}.\\ (b) Generate a large number of Gaussian random numbers employing Ferna\'ndez-Criado algorithm and plot their frequency distribution. Compare with the exact Gaussian.

\end{boxedtext}

\begin{figure}[ht]
\centerline{\psfig{figure=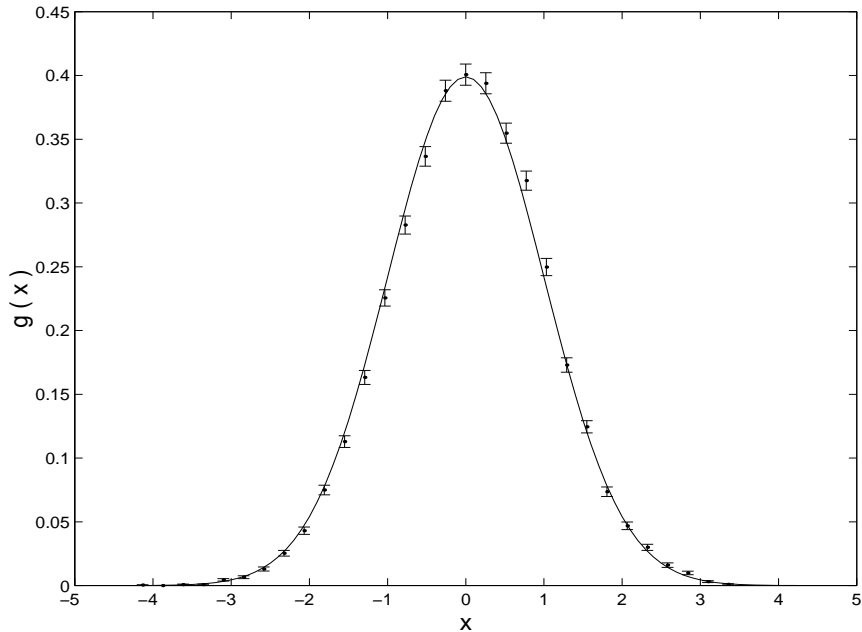,height=8.5cm,width=11.5cm}}
\caption{\small\protect Sampling from a Gaussian of mean zero and variance unity employing the technique proposed by Ferna\'ndez and Criado [39]. 
  $10,000$ particles were considered; warm up time was $40,000$ iterations; $20,000$ random numbers were generated and assembled in thirty equal bins; the error bar corresponds to one-sigma confidence in each bin. The solid curve is the exact Gaussian}
\label{fergauss_ps}
\end{figure}

We have implemented the algorithm proposed by Ferna\'ndez-Criado and a sample 
result is depicted in  Fig. \ref{fergauss_ps}.  Ten thousand 
particles were considered. The first $40,000$ iterations were considered 
as warm up time. We generated twenty thousand numbers and collected them 
in thirty equal bins. The statistical error (one - sigma confidence interval) 
on the counts in each bin were also calculated  from the sample fluctuations. 
These are also  depicted in  Fig. \ref{fergauss_ps}.

\noindent
\subsection{Metropolis Sampling }

\noindent
An interesting technique to sample from a probability distribution is the one proposed by Metropolis and his collaborators \cite{metropolis}, in the context of generating microstates belonging to a canonical ensemble. {\bf  Metropolis  algorithm} is widely used in Monte Carlo simulation of models in statistical physics.

Here I shall illustrate the technique  for sampling from an arbitrary discrete 
distribution $f(i)$, where the random variable takes discrete integer  
values $i$ between $1$ and $N$. We call $1,2,\cdots ,N$ as states and denote 
the set of all states by $\Omega=\{ 1,2,\cdots N\}$.  Start with an initial 
arbitrary state $x_0$ belonging to $\Omega$. The Metropolis sampling technique 
generates a {\bf Markov chain} of states $x_1 (\in \Omega), x_2 (\in \Omega) , 
\cdots x_{m-1}(\in \Omega)$.  For $m\to\infty$, 
$\{ x_{m+1}(\in\Omega),x_{m+2}(\in\Omega),\cdots\}$ 
shall have the distribution $ \{ f(i):i=1,2,\cdots ,N\}$. A Markov  chain 
is a stochastic process 
whose\  \ \lq past\rq\   has no influence on the future 
if its\  \lq present\rq\   is specified. In
other words, for a Markov Chain, we have, 
\begin{equation}
P\left( x_k \le x\vert x_{k-1},x_{k-2},\cdots , x_0\right) = P\left( x_k
\le x\vert x_{k-1}\right).
\end{equation}
We call $\{ x_{m+1},x_{m+2},\cdots\} $ the desired ensemble of states. 
If $N_i$ is number of elements in the ensemble whose state index is $i$, 
then $N_i$ divided by the total number of members in the ensemble is the 
probability $f(i)$. For such  asymptotic convergence to the desired 
ensemble, it is sufficient {\it though not   necessary} that
\begin{equation}\label{ONSAGER_EQ}
f(x_i)\ W(x_j \leftarrow x_i)=f(x_j)\ W(x_i\leftarrow x_j ) .
\end{equation}
where $W(x_j\leftarrow x_i)$ is the probability of transition from state $x_i$ to state $x_j$. Equation (\ref{ONSAGER_EQ}) is called the {\bf detailed balance}. For conservation of probability, the transition probabilities should obey the following condition,
\begin{equation}
\sum_{i} W(i \leftarrow j)=1\ \forall\ j . 
\end{equation}
The detailed balance condition does not specify uniquely the transition probabilities from one state to the other. A simple choice of the $N\times N$ {\bf transition  matrix} $W$ with elements $W_{i,j}=W(i\leftarrow j)$ which is consistent with the detailed balance is given by,
\begin{eqnarray}\label{WMATRIX_EQ}
W_{i,j} & =& W^{\star}_{i,j}\ \ {\rm min}\ \left(
1,{{f(i)}\over{f(j)}}\right)
\ {\rm for}
                     \ \   i\ne j,\nonumber\\ W_{j,j} &=& W^{\star}_{j,j}
+ \sum_{i=1}^{N} W^{\star}_{i,j}\ \ {\rm max}\ \left( 0,1-{{f(i)}\over{
f(j)}}\right),
\end{eqnarray}
where the matrix  $W^{\star}$ with elements $W^{\star}_{i,j}$ is an arbitrary symmetric stochastic matrix with positive elements.  We call $W^{\star}$ the {\bf trial } matrix. The sum of the elements in each row as well as each column of the trial matrix $W^{\star}$ is unity.  As we shall see below, we need $W^{\star}$ matrix to select a trial state from the current state. Hence $W^{\star}$ is chosen conveniently to make the selection of the trial state simple.

\begin{boxedtext}
\noindent
{\bf Assignment 18}\\

\noindent
Verify that the transition matrix $W$ whose elements are calculated as per the prescriptions in Eqns. (\ref{WMATRIX_EQ}) obeys the detailed balance given by Eq. (\ref{ONSAGER_EQ} ).
\end{boxedtext}

The implementation of the Metropolis sampling procedure proceeds as follows.  Let $x_i(\in \Omega)$ be the current state (or value).  We select a trial  state $x_t$,  randomly and with equal probability from amongst  the $N$ states of $\Omega=
\{ 1,2,\cdots N\}$.  This in other words means that $W_{i,j}^{\star}=1/N\ \forall \ i,j =1,2,\cdots ,N$. Calculate the ratio $w=f(x_t)/f(x_i)$. If $w\ge 1$, accept the trial state
and set $x_{i+1} = x_t $. If $w < 1 $, then  generate a random number $\xi $ uniformly distributed in the range $0$ to $1$.  If $\xi \le w $, then accept the trial state  and set $x_{i+1} =x_t$. If $\xi > w $, reject the trial  state  and set $x_{i+1} = x_i $ and proceed. It may be necessary to generate several values of $x$, starting from an initial choice $x_0$, before the string acquires the desired distribution. A good choice of $x_0$ is that state  for which the probability is maximum. 

Let us consider a simple example with three states $\Omega = \{ 1,2,3\}$ with $\vert f\rangle =(f_1,f_2,f_3)'=(1/4,\  5/12,\  1/3)'$.  The matrix $W^{\star}$ is taken as $W^{\star}_{i,j}=1/3\ \forall\quad i,j$.  The transition matrix constructed by the prescription, see Eq. (\ref{WMATRIX_EQ}), is given by,
\begin{equation}
W=\left( \begin{array}{ccc} 1/3 & 1/5  & 1/4\\ 1/3 & 8/15 & 1/3\\ 1/3 &
4/15 & 5/12 \end{array}\right).
\end{equation}
As can be readily seen, the matrix $W$ above, has the following properties.
\begin{enumerate}
\item
 $W_{i,j}> 0\  \forall\  i,j$; This  is called the {\bf strong ergodicity} condition.
\item
 $\sum_{i=1}^3 W_{i,j} =1\ \  \forall\ \  j$. As mentioned earlier, this condition ensures the conservation of the probability.  The transpose of the matrix $W$ is usually called  a stochastic matrix.
\end{enumerate}
 The eigenvalues of $W$ are $1.0,\ 0.2,\ {\rm and}\ 0.0833$; the largest eigenvalue is  unity and is non degenerate. The corresponding right eigenvector is $\vert f\rangle=(1/4,\ 5/12,\ 1/3)'$ {\i.e.,} $W\vert f\rangle = \vert f\rangle$.

\begin{boxedtext}
\noindent
{\bf Assignment 19}\\

\noindent
Construct a matrix $W$: $W_{i,j} > 0\ \forall\ i,j;$  and $\sum_{i} W_{i,j} = 1 \forall \  j$. Calculate its eigenvalues and the corresponding left and right eigenvectors.  Operate $W$ on an arbitrary vector several times and show that the resulting vector converges to a unique vector which is the eigenvector of the matrix $W$, corresponding the largest eigenvalue unity.
\end{boxedtext}

Thus repeated application of $W$ on any arbitrary vector $\vert u\rangle$ with 
$\langle f\vert u \rangle\ne 0$, will asymptotically take the vector to  $\vert f\rangle$. 
We say $\vert f\rangle$ is the equilibrium probability vector representing the equilibrium ensemble. 
Any initial ensemble represented by $\vert u\rangle$ with non-zero overlap with the 
equilibrium ensemble, will evolve to the equilibrium ensemble.  The above results 
are true in general for any  positive stochastic matrix $W$, by virtue of 
the {\bf Peron theorem} \cite{Peron}.

Perons's theorem states that {\it a positive matrix $W$ has an eigenvalue 
$\lambda$, which is real, positive and non-degenerate and which exceeds in modulus 
all the other eigenvalues.  To this dominant eigenvalue there corresponds an eigenvector 
with all its elements positive.} In our example, the transpose of the positive matrix $W$ 
is  stochastic, {\it i.e.,} $\sum_i W_{i,j}=1$, and hence the dominant eigenvalue is unity.

Peron's theorem can be seen as follows. Consider the following left eigenvalue equation,
\begin{equation}\label{LEIG_EQ}
\langle L\vert\  W = \langle L \vert\  \lambda\ ,
\end{equation}
where $\langle L\vert$ is the left eigenvector corresponding to the
eigenvalue $\lambda$.  We can obtain the upper and lower bounds of the
eigenvalue as follows.
\begin{eqnarray}\label{BOUND_EQ}
\lambda L_j & = & \sum_{i=1}^{N} W_{i,j}L_i\ ; \ \  \\
L_{min}\sum_{i=1}^N W_{i,j} \le \lambda L_j & \le &
L_{max}\sum_{i=1}^{N}W_{i,j},
\end{eqnarray}
where $L_{max}=max\{ L_k\}$ and $L_{min}=min\{ L_k \}$.  Since
$\sum_{i=1}^{N} W_{i,j} =1$, it follows,
\begin{equation}\label{BOUND2_EQ}
 L_{min}  \le \lambda L_j \le L_{max} \ .
\end{equation}
Consider the space of all positive vectors $\left\{ \langle L\vert ; L_j > 0 \ \forall\ j=1,2,\cdots ,N\right\}$. Then
\begin{equation}
{{L_{min} }\over{L_j }} \ \le\  \lambda\  \le\  {{L_{max} }\over{L_j }}\ \
\forall\ j.
\end{equation}
The minimum  value that $L_{max}/L_j$ can take as $j$ runs from $1$ to $N$ is unity;  similarly, the maximum value that $L_{min}/L_j$ can take is unity. Thus we get $1\le\lambda\le 1$, which implies that $\lambda =1$. Thus if the left eigenvector is positive then the corresponding eigenvalue is unity.

Consider now the space of vectors $\{ \langle L\vert\}$ such that each vector has  some of its components positive and some negative. We note that $L_{max}>0$ and $L_{min}<0$.  The bounds can be easily worked out and are given by,

\begin{equation}
{\rm max}\left\{ \ {{L_{ {\rm max}} }\over{ L_{{\rm min}} }}\ ,\ {{L_{{\rm
min}} }\over{ L_{{\rm max}} }}\ \right\}\
\le \ \lambda\  \le\  1.
\end{equation}

\noindent
Thus we see that $\lambda =1$ is the largest eigenvalue and the corresponding left eigenvector is positive.  It is easily verified that $\langle U\vert = c\ (1\  1\  1\ \cdots \ 1)$, the constant vector,  is the left eigenvector corresponding to the eigenvalue $\lambda =1$. The constant $c$ can be fixed by choosing a suitable norm.

We can easily show that the eigenvalue $\lambda =1$ is non degenerate.  To this end, assume the contrary. Let $\langle V\vert$ be another positive eigenvector corresponding to the eigenvalue $\lambda =1$. Then a linear combination of $\langle U\vert$ and $\langle V\vert$ given by $\langle \eta\vert=\alpha \langle U\vert + \beta\langle V\vert$ is also an eigenvector.  We can choose $\alpha$ and $\beta$ such that $\langle \eta\vert$ has some components positive and some negative which contradicts the fact that the eigenvector is positive. This completes the proof.

Let us now consider  the right eigenvector $\vert R\rangle$, corresponding to  $\lambda =1$.  It is easily proved that $\vert R\rangle$ is positive. Consider the eigenvalue equation 
\begin{equation}\label{righteigv_eq} (W_{1,1} -1)R_1 + W_{1,2}R_2 +\cdots +W_{1,N} R_N =0.
\end{equation}
Let us assume $R_1$ is negative. The first term in Eq. (\ref{righteigv_eq}) is positive since $W_{1,1} < 1$. Hence  at least one of the other components of $\vert R\rangle$ must be negative. In other words one of the elements of the set $\{ R_2 ,\  R_3 ,\ \cdots R_N \}$ must be negative to render the sum in Eq. (\ref{righteigv_eq}) zero. Without loss of generality we take $R_2$ to be negative. Consider the eigenvalue equation,
\begin{equation}\label{righteigv2_eq}
W_{2,1}R_1 + (W_{2,2}-1)R_2 +W_{2,3}R_3 + \cdots + W_{2,N}R_N =0.
\end{equation}
Add Eq. (\ref{righteigv_eq}) to Eq. (\ref{righteigv2_eq}) and get,
\begin{equation}\label{sumonetwo}
(W_{1,1} + W_{2,1}-1)R_1 + (W_{1,2} + W_{2,2} -1)R_2 + \sum_{j=3}^{N}
(W_{1,j}+W_{2,j})R_j =0.
\end{equation}
In the above, $(W_{1,1}+W_{2,1})\ <\ 1$, $R_1 < 0$, $(W_{1,2} + W_{2,2}) <  1$ and $R_2 < 0$.  Therefore we find that the first two terms in the above equation are positive. Hence at least one of the elements of the set $\{ R_3 ,\ R_4 \  \cdots R_N \}$ must be negative  to render the sum zero. Without loss of generality we take this as $R_3$.  Arguing along the same lines we show that if $R_1$ is negative then all the other components of the vector $\vert R\rangle$ are also negative. This in other words means that the right eigenvector corresponding to the largest eigenvalue $\lambda =1$ is positive. We call $\vert R\rangle$ the equilibrium eigenvector.
  
Numerically the equilibrium ensemble for the three-state model is constructed as follows. Start with an arbitrary state $x_0\in\Omega$; select a trial state $x_t\in\Omega$ with equal probability. Note that we have taken $W^{\star}_{i,j}=1/3\ \forall\ i,j$.  Calculate $w=f(x_t)/f(x_0)$.  Accept the trial state as the next state with a probability given by minimum of $(1,w)$. If  accepted set $x_1=x_t$; if
rejected set $x_1=x_0$; and proceed. Thus we get an ensemble of states.
  
Peron's  theorem is special case of a more general theorem on the eigenvectors and eigenvalues of non-negative matrices proved later by Frobenius. A proof of the  {\bf Frobenius theorem} can be found in the book on matrices by Gantmacher \cite{GANTMACHER}. As per this theorem, asymptotic convergence to equilibrium vector is possible even under weaker condition. Some or several elements of the transition matrix $W$ can be zero.  In other words $W_{i,j} \ge 0\ \forall i,j$ and of course $\sum_I W_{i,j}= 1$. In such cases it is adequate that $\left( W^m \right)_{i,j}>
0\  \forall \ i,j $ and for an $m< \infty$; $m$ may depend on $i$ and $j$. Physically this means that any state is accessible from any other state in finite number of steps.

This weaker condition helps us make a better choice of the trial matrix $W^{\star}$.  Some or several elements of this matrix can be zero. In practical terms it means that the trial state $x_t$ can be selected randomly from a small neighbourhood of the current state $x_i$.  Thus we set $x_t = x_i + \eta_i $ where $\eta_i$ is a random integer from $-\epsilon $ to $+\epsilon$.  In this approach, it may so happen that the trial value lies outside the range of $x$. In such a situation, keep generating new trial values until, you get one within the range.  The choice of $\epsilon$ is crucial. If $\epsilon $ is too large, then the fraction of the accepted trials would  be too small and the sampling poor and inefficient. On the other hand if $\epsilon $ is too small, then even though a large number of trials would  get accepted, the value of $x$ would  remain close to the starting point over several trials and hence
not span the entire range of $x$ quickly and efficiently.  A good criterion is to fix $\epsilon$ such that the acceptance is between $30\%$ and $50\%$.

\begin{boxedtext}
\noindent
{\bf Assignment 20}\\

\noindent
Employ Metropolis sampling technique to sample from a) binomial distribution and b) Poisson distribution.
\end{boxedtext} 

\noindent
In the above we have discussed the Metropolis technique with respect to discrete distributions. The method can be readily extended to sample from continuous distribution.
\begin{figure}[ht]
\centerline{\psfig{figure=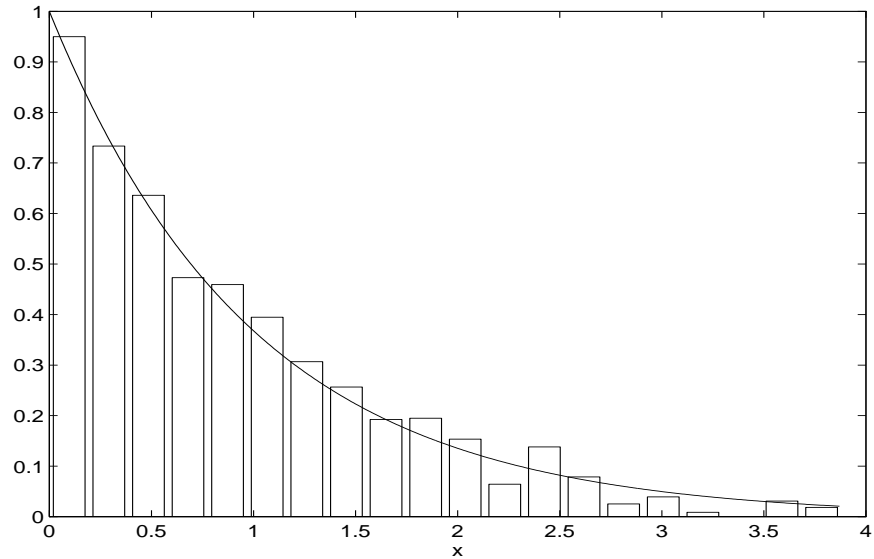,height=7.5cm,width=11.5cm
}}
\caption{\small\protect Sampling from an exponential density employing Metropolis algorithm. $\epsilon =3$ gives an acceptance of $30\% - 50\%$. The solid curve is the exact exponential distribution. The histogram represents the frequency distribution of a sample of size $10,000$.  }
\label{METROPOLIS_PS}
\end{figure}

Fig. \ref{METROPOLIS_PS} depicts the results of sampling from an exponential density employing Metropolis sampling.

A very large number of transformations, tricks, and formulae have been discovered for sampling from a variety of non-uniform distributions. I shall not discuss them here. These techniques are scattered over several publications on Monte Carlo and its applications to different disciplines.

\begin{boxedtext}
\noindent
{\bf Assignment 21}\\

\noindent
Employ Metropolis technique to sample from the distribution $f(\theta)=\exp\left[ a\cos(\theta)\right]$, for $0\le \theta \le \pi$, where $a$ is a parameter. The  optimal value of $\epsilon$ that would give $30\%$ to $50\%$ rejection would depend  on the value of  $a$.  Compare your results with the exact.
\end{boxedtext}

\noindent
\section{ANALOGUE MONTE CARLO}

\noindent
Consider evaluation of the expectation value (also known as the mean) of a function $h$ of the random variable $X$. Formally we have, 
\begin{equation}
\mu=\int_{-\infty}^{+\infty} h(x)f(x)dx,
\end{equation}
where $f(x)$ is the probability density function of the random variable $X$. $h(x)$ is usually called the score function. How do we estimate $\mu$~?  Let us first consider {\bf analogue simulation}. A Monte Carlo simulation is called analogue simulation, if it does not employ any variance reduction devices. The simulations that employ variance reduction techniques we shall term as biased simulation.  In analogue Monte Carlo, we sample randomly a sequence of $\{ x_i\  :i=1,2,\cdots ,N\}$, from the density $f(x)$ and write,
\begin{equation}
\bar{h}_N = {{1}\over{N}}\sum_{i=1}^{N} h(x_i) .
\end{equation}
In the limit $N\to\infty$, $\bar{h}_N \to\mu$. Also by the Central Limit Theorem, in the limit $N\to\infty$, the probability density of $\bar{h}_N$ tends to a Gaussian with mean $\mu$ and variance $\sigma^2 /N$, where
\begin{equation}
\sigma^2 = \int_{-\infty}^{+\infty}\left[ h(x)-\mu\right]^2f(x)dx .
\end{equation}
Thus we say that {\bf Analogue Monte Carlo} estimate of $\mu$ is given by $\bar{h}_N \pm \sigma/\sqrt{N}$, where $\pm\sigma/\sqrt{N}$ defines the one-sigma confidence interval. This means that with a probability $p$ given by,
\begin{eqnarray}
p &=& {{\sqrt{N} }\over{\sigma\sqrt{2\pi} } }
\int_{\mu-(\sigma/\sqrt{N})}^{\mu +(\sigma/\sqrt{N})} \exp\left[
-{{N(x-\mu)^2}\over{2\sigma^2}}\right] dx\nonumber\\
\ &=& {{1}\over{\sqrt{2\pi}}} \int_{-1}^{+1}\exp
           \left[ -{{x^2}\over{2}}\right] dx\nonumber\\
\ &=& 0.68268 ,
\end{eqnarray}
we expect $\bar{h}_N$ to lie within $\pm \sigma/\sqrt{N}$ around $\mu$, if $N$ is sufficiently large.

First we note that we do not know $\sigma$. Hence we approximate it by its Monte Carlo estimate $S_N$, given by
\begin{equation}
S^2 _N = {{1}\over{N}}\sum_{i=1}^{N}h^2 (x_i ) -\left[ {{1}\over{N}}
\sum_{i=1}^{N} h(x_i )\right] ^2 .
\end{equation}
The quantity $\pm S_N /\sqrt{N}$ is called the {\bf statistical error}. Notice that the sample size $N$ must be large for the above estimate of the error (and of course the mean) to hold well.  Sometimes it would worth the effort to test if the sample mean has acquired a normal distribution, see for example the test devised by Shapiro and Wilks \cite{SHAP}. Normality tests are useful in a biased Monte Carlo simulation.

The statistical error decreases as inverse of the square root of the sample size. This is rather slow - logarithmically slow. For example if we want to decrease the statistical error by a factor of two we must increase the  sample size by a factor of four. The computing time increases linearly with the sample size. Hence invariably one finds that analogue Monte Carlo is practically impossible.  But there is a way out.

The way out is to resort to techniques that reduce the variance 
without altering the mean. These are called 
{\bf variance reduction techniques} and in what follows I shall 
describe the basic principle behind variance reduction techniques 
through what is called the {\bf importance sampling}.

\noindent
\section{IMPORTANCE SAMPLING }

\noindent
Importance sampling helps us sample from the important regions of the sample space. Consider the problem described in the last section.  We sample $\{ x_i :i=1,2,\cdots ,N\}$ from an importance density $g(x)$ instead of the analogue density $f(x)$. To preserve the mean we define a modified score function $H(x)$ given by,
\begin{equation}
H(x)=h(x){{f(x)}\over{g(x)}}.
\end{equation}
The expectation value of $H$ is evaluated over the importance density $g(x)$; this is 
identically equal to the expectation value of the original score function $h$ over the analogue density $f(x)$:
\begin{eqnarray}
\mu (H)&=&\int_{-\infty}^{+\infty} H(x)g(x) dx\nonumber\\
\      &=&\int_{-\infty}^{+\infty} h(x) {{f(x)}\over{g(x)}} g(x) dx\nonumber\\
\      &=&\int_{-\infty}^{+\infty} h(x)f(x)dx\nonumber\\
\      &=&\mu(h) .
\end{eqnarray}
Thus we sample $\{ x_i :i=1,2,\cdots ,N\}$ from $g(x)$, and calculate the ratios $\{
f(x_i)/g(x_i)\} $. The biased Monte Carlo estimate of $\mu$ is given by
\begin{equation}
\bar{H}_N={{1}\over{N}}\sum_{i=1}^{N}h(x_i ){{f(x_i )}\over{g(x_i )}} .
\end{equation}
In the limit $N\to\infty$, $\bar{H}_N\to\mu$. Let us now calculate the statistical error associated with $\bar{H}_N$. It is adequate if we consider the second moment, since we have formally shown that the mean is preserved under importance sampling. We have
\begin{eqnarray}\label{m2bh}
M_{2}^{B} (H) &=& \int_{-\infty}^{+\infty} H^2 (x)g(x) dx\nonumber\\
\             &=& \int_{-\infty}^{+\infty}
                   {{h(x) f(x)}\over{g(x)}} {{h(x)f(x)}\over{g(x)}} g(x)
dx\nonumber\\
\             &=&\int_{-\infty}^{+\infty}
                    \left[ {{f(x)}\over{g(x)}}\right]  h^2 (x) f(x) dx .
\end{eqnarray}
For the analogue simulation,
\begin{equation}
M_{2}^{A} (h)=\int_{-\infty}^{+\infty} h^2 (x)f(x) dx .
\end{equation}
Thus if we choose properly the importance function $g(x)$ and ensure that the ratio $f(x)/g(x)$ on the average is substantially less than unity, then we can make $M_{2}^{B} (H)$ to be much less than $M_{2}^{A}(h)$. This in essence is the basic principle of variance reduction techniques.  Thus, sampling from an importance density helps us estimate the mean with a much better statistical accuracy for a given sample size. In fact it is due to the variance reduction techniques that Monte Carlo simulation of many problems have become possible.  

\noindent
\subsection{Exponential Biasing}

\noindent
Let me illustrate the  use of importance function on a simple problem where all the relevant quantities can be calculated analytically. The problem has its origin in radiation transport through thick shields.  The actual problem of Monte Carlo simulation of the transport is relatively complex and a detailed discussion of this is not relevant for the purpose here. Hence, I shall be brief.  Consider particles, neutrons or gammas, incident normally on the left face of a slab of  material of thickness $T$. The particle would penetrate a distance $x$ into the shield with a probability $\Sigma \exp (-\Sigma x)$ and enter into a collision event at a point between $x$ and $x+dx$ with probability $\Sigma dx$.  Here $\Sigma^{-1}$ is the mean free path (mfp). We can measure all distances in units of mfp and hence set $\Sigma =1$. The collision can lead to absorption in which case the particle history is
terminated.  On the other hand if the collision is a scattering event, the history continues. In a simple-minded picture, the scattering is taken as isotropic. In one-dimensional model, this means that the particle is directed toward the right or the left face of the slab, with equal probability.  The particle travels along the scattered direction and has a collision event at some distance. The history continues and is constituted by a series of alternating free flights and collisions. The history ends when the particle is absorbed inside the shield or escapes the shield.
When the particle escapes through the right face it scores unity; otherwise the score is zero. The scores from a large number of particle histories are accumulated. The average of these scores is the Monte Carlo estimate of the mean transmission defined as the fraction of the incident particles that escape through right face of the shield. The sample fluctuations give an estimate of the statistical error. If the thickness of the shield exceeds $20$ mfp or so, the problem becomes intractable by
analogue Monte Carlo. Variance reduction techniques become imperative.
 
An importance sampling technique often used in this context is called  the 
{\bf exponential biasing}. In the simplest version of exponential biasing, the inter collision distance is sampled from the importance density $b\exp(-bx)$, where the biasing parameter $b$ has to be optimized for minimum variance. For more details on exponential biasing see \cite{clark}.

In the simple problem, we shall be interested in calculating the fraction 
of the incident particles that escape through the right
face of the shield, without  
undergoing any collision whatsoever. This amounts to the 
following: In analogue simulation, sample a random number from the exponential 
density. If this exceeds $T$, score unity. Otherwise score zero. Collect the scores 
from $N$ histories and calculate the mean and the statistical error.  
In the biased simulation we sample $x$ from the importance density 
function $\hat{b}\exp(-\hat{b}x)$, where we have
assumed that we know the value  of $b=\hat{b}$, for which the variance is minimum. 
Let us denote this sampled value of $x$ as $x_i$. If $x_i$ exceeds $T$, then  
score $w_i=f(x_i)/g(x_i , \hat{b})= \exp [ -x_i (1-\hat{b})]/\hat{b}$. 
Otherwise  score
zero. Accumulate the scores from a large number of 
histories and calculate  the mean and the statistical error.

\begin{figure}[ht]
\centerline{\psfig{figure=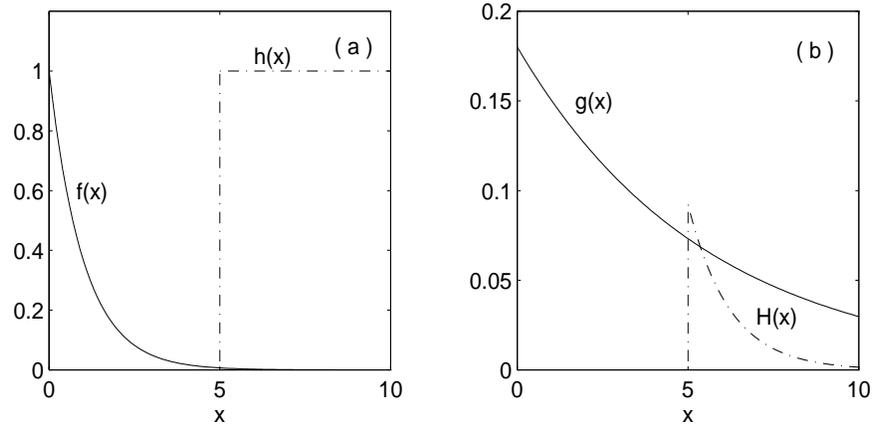,height=5.8cm,width=11.5cm}}
\caption{\protect\small  (a) The probability density $f(x)$ and the score 
function $h(x)$ for $T = 5$ mfp; (b) The importance density function
$g(x)$  and the modified score function $H(x)=
h(x)f(x)/g(x)$. $\hat{b}=0.18$}
\label{IMP_PS}
\end{figure}

Fig. \ref{IMP_PS}a depicts the probability density function $f(x)$ and 
the score function $h(x)$ for the case with $T=5$ mfp.  
Fig. \ref{IMP_PS}b depicts the importance density function 
$g(x,b=\hat{b})$ and the modified score function 
$H(x)=f(x)h(x)/g(x,\hat{b})$ for the case with $T=5$ mfp.

For this problem, all the quantities can be calculated analytically.  
The expressions for the average and variance are,
\begin{eqnarray}
\mu  &=&  \int _0 ^{\infty} h(x)\exp (-x)\ dx=\exp(-T) ,\\
\sigma^2 & = &  ( 1-\mu)\ \mu.
\end{eqnarray}
Table (6) 
gives the mean, relative standard deviation and the number of histories 
that would be required to estimate the mean in an analogue simulation, 
within $\pm 10\%$ of the exact.

\begin{table}[h]\label{anasimexact_tab} 
\caption{\protect\small  Analogue simulation: exact analytical results}
\bigskip
\begin{center}
\begin{tabular}{|c|c|c|c|}
\hline  
     &                      &                   &                     \\
$T$  & $\mu$                & $\sigma/\mu$      & $N$ \\ &
&                   &                     \\
\hline
   &                        &                      &  \\ $3$  &
$4.98\times 10^{-2}$ & $4.37$            & $1.91\times 10^3   $\\ $5$  &
$6.74\times 10^{-3}$ & $1.21\times 10^1$ & $1.47\times 10^4   $\\ $10$ &
$4.54\times 10^{-5}$ & $1.48\times 10^2$ & $2.20\times 10^6   $\\ $20$ &
$2.06\times 10^{-9}$ & $2.20\times 10^4$ & $4.85\times 10^{10}$\\ &
&                   &                     \\
\hline
\end{tabular}
\end{center}
\end{table}

We see from Table (6) 
that as the thickness increases the mean transmission decreases 
and the relative fluctuation increases. We find that the number of 
histories required to predict the mean within $\pm 10\%$ statistical 
error is over $48$ billion for $T=20$ mfp, a task which is clearly 
impossible on any computer.

Let us see how the use of importance sampling renders possible 
the impossible. Under importance sampling,  the variance 
can be calculated analytically and is given by,
\begin{equation}
\sigma ^2 (b)= {{ e^{-T(2-b)} }\over{b(2-b)}}-e^{-2T}.
\end{equation}
It is quite straight forward to calculate the value of $\hat{b}$ for which
the variance is minimum. We get,
\begin{equation}
\hat{b} = 1+ {{1}\over{T}}-\sqrt{1+{{1}\over{T^2}} }.
\end{equation}
Table (7) 
presents $T, \hat{b}, \sigma /\mu$ and the number of histories 
required to estimate the mean, under biased simulation  within $\pm 10\%$ of the exact.

\begin{table}[h]\label{bse_tab}
\caption{\protect\small  Biased Simulation: exact analytical results}
\bigskip
\begin{center}
\begin{tabular}{|c|c|c|c|}
\hline
  &         &               &  \\
\ \ $T$\ \  & \ \ $\hat{b}$\ \  &\ \  $\sigma / \mu$ \ \  &\ \  $N$\ \ \\
  &         &               &  \\
\hline
    &           &                 &      \\ $\ 3$ & $.28$     & $1.95$
& $\ 381$\\ $\ 5$ & $.18$ & $2.55$ & $\ 651$\\ $10$ & $.09$ & $3.65$ &
$1329$\\ $20$ & $.048$& $5.18$ & $2687$ \\ &        &        &       \\
\hline
\end{tabular}
\end{center}
\end{table}

We find from Table (7) 
that the use of importance sampling would lead to a 
considerable reduction of variance 
especially for large $T$. Take the case with $T=5$ mfp. 
The use of importance 
sampling would reduce the statistical 
error by a factor five or so.   As a consequence 
a sample of size $650$ is adequate to estimate the 
mean whereas analogue simulation 
would require a sample of size $14,700$.  
The results for $T=20$ mfp are more dramatic 
and bring home the need and power of importance sampling. The statistical 
error gets reduced by a factor of
$4247$.  We need to simulate only $2687$ histories to estimate the mean 
within $\pm 10\%$ of the exact, as compared to $48$ billion histories required 
under analogue simulation.

We have simulated explicitly $10000$ histories under importance sampling, as follows. 
We sample $x_i$ from the importance density, $\hat{b}\exp (-\hat{b}x)$, and 
calculate the mean of the modified score function as, 
\begin{equation}
\bar{H}_N = {{1}\over{N}}\sum_{i=1}^N H(x_i),
\end{equation}
where,
\begin{equation}
H(x_i)=\left\{ \begin{array}{lll} {{1}\over{\hat{b}}}\exp \left[ -\left(
1-\hat{b}\right) x_i\right] , \ & {\rm if }& x_i\ \ge \ T,\\ &
\\ 0,                    & {\rm if} & x_i\  < \  T.  \end{array}\right.
\end{equation}
The statistical error is calculated as,
\begin{equation}
\Delta \bar{H}_N = \pm {{1}\over{\sqrt{N}}}\sqrt{
                      {{1}\over{N}}\sum_{i=1}^{N}H^2 (x_i)-\bar{H}_N ^2 }.
\end{equation} 
Table (8) 
gives the estimated mean
transmission $\bar{H}_N$, the relative statistical error, and the actual
deviation of the Monte Carlo estimate from the exact transmission.

\begin{table}[h]\label{tenthousand_tab}
\caption{\protect\small  Results of Monte Carlo simulation of $10,000$ histories
with importance sampling  and comparison with exact analytical
calculations}
\bigskip
\begin{center}
\begin{tabular}{|c|c|c|c|}
\hline
   &        &     & \\ $T$&$\bar{H}_N$ & ${{\Delta
\bar{H}_N}\over{\bar{H}_N}}\times 100$ & $ {{\bar{H}_N
-\mu}\over{\mu}}\times 100$\\ &                             &
&       \\
\hline
   &                                  &                & \\ $3$   &
$4.94\times 10^{-2}$         &   $\pm 2.0\%$  &$ -0.8\%$\\ $5$   &
$6.76\times 10^{-3}$         &   $\pm 2.6\%$  &$ +0.3\%$ \\ $10$  &
$4.52\times 10^{-5}$         &   $\pm 3.7\%$  & $-0.4\%$\\ $20$  &
$1.96\times 10^{-9}$         &   $\pm 5.4\%$  & $-4.9\%$\\ &
&                & \\
\hline
\end{tabular}
\end{center}
\end{table}

We observe from Table (8) 
that we are able to make a good estimate of the mean transmission employing importance sampling. 
The corresponding results obtained by analogue simulation of $50000$ histories (five times more than what we have 
considered for analogue simulation) 
are given in Table (9).  

\begin{table}[h]\label{fiftythousand_tab}
\caption{\protect\small  Results of analogue simulation of $50,000 $ histories and
comparison with exact analytical results}
\bigskip
\begin{center}
\begin{tabular}{|c|c|c|c|}
\hline
    &              &                    &\\ $T$ & $\bar{h}_N$ & ${{ \Delta
\bar{h}_N}\over{\mu}}\times 100$ & ${{\bar{h}_N-\mu}\over{\mu}}\times
100$\\ &    &                                           &         \\
\hline
    &                        &                 &          \\ $3$  &
$5.03\times 10^{-2}$   &   $\pm 1.9\%$   &$+ 1.8\% $ \\ $5$  &$6.68\times
10^{-3}$   &   $\pm 5.5\%$   &$-0.9\%$  \\ $10$ &$6.16\times 10^{-4}$   &
$\pm 57.7\%$  &$+35.7\%$  \\ $20$ &$\cdot$                 &   $\cdot$
&$\cdot$  \\ &                       &                  &\\
\hline   
\end{tabular}
\end{center}
\end{table}
 
We find from Table (9) 
that analogue simulation of the problem with $T=20$ mfp is impossible.  
On the average, we can expect one in $49$ million numbers sampled from 
the exponential density, to have a value greater than $20$. 
The chance of getting a score in a simulation of $50000$ histories is practically nil.  

\noindent
\subsection{Spanier  Technique  }

\noindent
In several situations, like the one discussed in section 12.1, it is possible 
to make a good guess of the shape of the importance function $g(x,b)$, where 
$b$ is an unknown parameter to be optimized for minimum variance. 
Spanier \cite{spanier} has proposed a technique for optimizing the parameter $b$, 
through processing of the Monte Carlo results from relatively  small samples 
of histories.  The technique consists of expressing the second moment, 
for the biased problem, see Eq. (\ref{m2bh}) as,
\begin{equation}
M_2 (b_i) = \int {{ h(x)f(x)}\over{g(x,b_i)}}\ {{
h(x)f(x)}\over{g(x,\tilde{b})}} \  g(x,\tilde{b})dx\ ,
\end{equation}
where $\tilde{b}$ is the guess value of the  parameter $b$ and $\{ b_i :
i=1,2,\cdots ,M\}$ are the prescribed values of $b$ spanning its full range or a
conveniently chosen interval in its range.

 The implementation of {\bf Spanier's technique} proceeds as follows. Start with a guess value $\tilde{b}$. Generate a set of $N$ values of $x$ by random sampling from the importance density $g(x,\tilde{b})$. Calculate,
\begin{equation}
M_2 (b_j) = {{1}\over{N}} \sum_{i=1}^{N} {{
h(x_i)f(x_i)}\over{g(x_i,b_j)}}\ {{ h(x_i)f(x_i)}\over{g(x_i,\tilde{b})}}
\quad {\rm for} \quad j=1,2,\cdots ,M.
\end{equation}
Thus we get $M_2$ at $M$ distinct values of $b$. Find the value of $b$ at which the second moment is minimum;  use this as $\tilde{b}$ in the next iteration and generate a second set of $N$ histories. Repeat the procedure until $\tilde{b}$ converges. $N$ can be small for the purpose of optimizing $b$.

\begin{figure}[ht]
\centerline{\psfig{figure=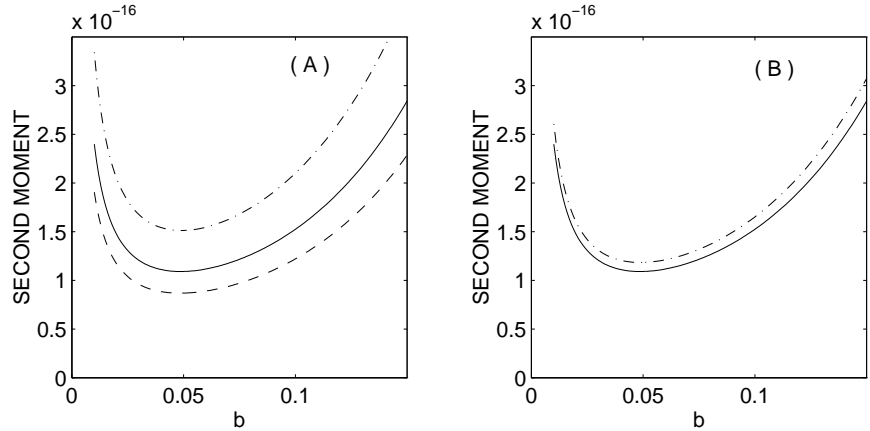,height=5.8cm,width=11.5cm}}
\caption{\protect\small (A)\  \lq  - - \rq\   denotes the results generated by
$\tilde{b}=0.1$; the minimum of the curve occurs at $b=0.048$;
\ \lq -$\cdot$-$\cdot$\rq\   
represents the results generated with $\tilde{b}=0.048$ and the minimum
occurs at $b=0.049$. \ \lq - \rq\   denotes the results generated with
$\tilde{b}=0.049$ and the minimum is found to be at $b=0.049$. $500$
histories were generated at each stage of iteration. (B)\ \lq -\rq\
depicts the second moment generated by the converged value
$\tilde{b}=0.049$ and
\ \lq -$\cdot$-$\cdot$\rq\    depicts the exact analytical results.}
\label{SPANIER_PS}
\end{figure}

Fig. \ref{SPANIER_PS}A presents the results for the 
problem with $T=20$ mfp at three stages of iteration 
starting with an initial guess 
of $\tilde{b}=0.1$. The number of histories generated 
at each stage of iteration is $500$.   Fig.  \ref{SPANIER_PS}B 
depicts the second moment as a function of $b$ generated 
with the converged value of $\tilde{b}=.049$, along with the exact analytical results.

Spanier's technique is quite general and powerful. Application of this technique to radiation transport problem can be found in \cite{spanier, murthy}. Several modifications and improvements  to this technique have been recommended and those interested can refer to \cite{mac}. 

\noindent
\section{CLOSING REMARKS }

\noindent
In this monograph I have made an attempt  to describe the fundamental aspects of Monte Carlo theory in very  general terms without  reference to any particular topics like neutron transport or problems in statistical mechanics.

An  understanding of the fundamental ideas of probability theory: sample space, events, probability of events, random variables, probability distribution, moments, cumulants, characteristic function, {\it etc}, is a must  for appreciating the basis of Monte Carlo techniques. I have given a brief introduction to these topics.

The Central Limit Theorem and its use in estimating Monte Carlo 
error bars is an important topic and has been dealt with in detail, 
preceded by a brief discussion on the Chebyshev inequality and the 
law of large numbers. I have also very briefly touched upon the 
generalization of the Central Limit Theorem, 
namely the L\'evy stable law.
 
A long sequence of pseudo random numbers constitutes the backbone of any Monte Carlo simulation.  Often one finds that many Monte Carlo practitioners tend to remain ignorant of how these numbers are produced. They often consider  the random number generator  routine in their personal computer,  the workstation  or the super computer, as a black box. This tendency must go. The reason is simple. There is no random number generator which is flawless and which would yield random numbers
useful  for all kinds of simulations. Take for example the most popular and widely used of the  random number  generators. It is based on linear congruential recursion; it  suffers from a very serious defect: the lattice and Marsaglia hyper plane structures that ensue when you embed the random numbers in two and higher dimensional phase space. If your problem is such that the Marsaglia lattice structure would influence your results significantly then you should seriously consider alternate means of generating pseudo random numbers.  I have given a brief outline of these
and related  issues  in this monograph.

I have also discussed some tests of randomness of the sequence of pseudo random numbers.  Of course we  understand that there is absolutely no guarantee that the  sequence of pseudo random numbers given by the generator in your machine  is adequately random for the particular Monte Carlo application you have on hand; there can at best  be only safeguards: carry out certain tests  of randomness, like the uniformity test, correlation tests,  run test,  the tests based on embedding the sequence in phase space through construction of delay  vectors or any other sophisticated tests you may think of.  In fact the Monte Carlo algorithm you have developed itself can be used to test the randomness of the pseudo random  number  sequence.
 
Random sampling from non-uniform distributions constitutes  the core 
of Monte Carlo simulations.  I have dealt with inversion, rejection, 
and Metropolis techniques for sampling from a given distribution. 
The last of these, namely the Metropolis algorithm, is 
usually employed in simulations of  problems in statistical mechanics.

Almost all Monte Carlo simulations employ, in some form or the other, techniques of variance reduction. I have tried  to convey the basic principles of variance reduction through a discussion  of the importance sampling device. I have illustrated the use of importance function by considering  exponential biasing on a simple problem.  I have also dealt briefly with the Spanier technique for optimizing the biasing parameter in importance sampling.

Let me  end by saying that there  are several excellent  books and review articles on Monte Carlo theory and practice, which you would like to consult during your \ \lq Monte Carlo learning\rq\  .  Some of these books that have caught my attention are listed under \cite{REF}. Let me wish you all a merry time with Monte Carlo games.

\newpage
\thispagestyle{empty}
\noindent
{\bf About the author......}\\

K. P. N. Murthy was born in Chennai and graduated from the Vivekananda College, Mylapore.  
He is presently a member of the Theoretical Studies Section of the Materials Science 
Division, Indira Gandhi Centre for Atomic Research, Kalpakkam.
 
K. P. N. Murthy obtained his master's degree in Physics from the University of Bombay 
for his thesis on Monte Carlo Radiation Transport. He did his doctoral work at the 
Central University of Hyderabad and obtained his Ph.~D.  in theoretical physics  for 
his thesis on fluctuation phenomenon in model non-equilibrium systems.  His fields of 
interest include random walks, Monte Carlo methods, stochastic processes, 
statistical physics, relaxation phenomena, radiation transport, regular and 
anomalous diffusion, disordered systems, time series analysis, 
and Chaos. 

\end{document}